\documentclass{article}

\usepackage[dblblindworkshop, final]{neurips_2026}

\usepackage[utf8]{inputenc} 
\usepackage[T1]{fontenc}    
\usepackage{hyperref}       
\usepackage{url}            
\usepackage{booktabs}       
\usepackage{amsfonts}       
\usepackage{nicefrac}       
\usepackage{microtype}      
\usepackage{xcolor}         

\usepackage{amsmath}
\usepackage{amssymb}
\usepackage{float}
\usepackage{algorithm}
\usepackage{algpseudocode}
\usepackage{caption}
\usepackage{multirow}
\usepackage{enumitem}
\setlist{nosep,leftmargin=*}
\usepackage{tikz}
\usetikzlibrary{positioning, arrows.meta, fit, backgrounds, calc}
\usepackage{listings}

\definecolor{pykw}{RGB}{0,90,180}
\definecolor{pycom}{RGB}{34,139,34}
\definecolor{pystr}{RGB}{170,55,55}
\definecolor{pyann}{RGB}{120,40,160}
\lstdefinestyle{pythonpolicy}{
  language=Python,
  basicstyle=\ttfamily\scriptsize,
  keywordstyle=\color{pykw}\bfseries,
  commentstyle=\color{pycom}\itshape,
  stringstyle=\color{pystr},
  showstringspaces=false,
  breaklines=true,
  breakatwhitespace=true,
  tabsize=2,
  numbers=left,
  numberstyle=\tiny\color{gray},
  numbersep=4pt,
  xleftmargin=14pt,
  framexleftmargin=4pt,
  frame=single,
  framerule=0.4pt,
  rulecolor=\color{black!30},
  belowskip=2pt,
  aboveskip=2pt,
  emph={waste_fraction,find_best_clean_orientation,bfs_toward,bfs_to_target_set,
        bfs_nearest_apple,bfs_to_waste,best_clean_orientation,count_waste_in_beam,
        zone_boundaries,_rotation_distance,direction_to_action,voronoi_zones,
        my_zone_apples,assign_apple_zones,bfs_nearest_apple_in_zone,
        get_waste_cells,get_my_apples,nearest_respawning_apple,get_waste_positions,
        deque,np},
  emphstyle=\color{pyann}
}

\lstdefinestyle{prompttext}{
  basicstyle=\ttfamily\scriptsize,
  showstringspaces=false,
  breaklines=true,
  breakatwhitespace=true,
  tabsize=2,
  numbers=none,
  xleftmargin=4pt,
  framexleftmargin=2pt,
  frame=single,
  framerule=0.4pt,
  rulecolor=\color{black!30},
  belowskip=2pt,
  aboveskip=2pt,
  columns=fullflexible,
  keepspaces=true,
  morecomment=[l][\color{pyann}\itshape]{\#\#},
  morecomment=[s][\color{pycom}\itshape]{[}{]},
  moredelim=[is][\color{pyann}\bfseries]{**}{**}
}

\definecolor{rescolor}{RGB}{245, 127, 32}
\definecolor{pipcolor}{RGB}{37, 99, 214}
\definecolor{ilcolor}{RGB}{132, 72, 212}
\definecolor{metcolor}{RGB}{6, 164, 120}
\definecolor{deccolor}{RGB}{214, 48, 64}
\definecolor{frzcolor}{RGB}{95, 107, 135}

\newcommand{\cR}{\mathcal{R}}
\newcommand{\cM}{\mathcal{M}}
\newcommand{\cG}{\mathcal{G}}
\newcommand{\cC}{\mathcal{C}}
\newcommand{\bm}{\mathbf{m}}

\title{Discovering Cooperative Pipelines: Autoresearch for Sequential Social Dilemmas}

\author{%
  Víctor Gallego \\
  Komorebi AI Technologies\\
  \texttt{victor.gallego@komorebi.ai} \\
}

\begin{document}

\maketitle

\begin{abstract}
  We study \emph{two-level autoresearch} for cooperation: an outer-loop AI agent autonomously redesigns the inner-loop pipeline of an LLM policy-synthesis system for multi-agent Sequential Social Dilemmas (SSDs). A \emph{researcher agent} $\cR$ (run as a coding agent) reads the inner-loop source code, edits system prompts, feedback functions, helper libraries, and iteration logic, runs evaluations, and decides what to keep, following the \emph{autoresearch} paradigm. Across two games (Cleanup and Gathering), two policy-synthesizer LLMs, and two welfare objectives (utilitarian efficiency and Rawlsian maximin), the researcher reliably exceeds hand-designed baselines, sharply tightens run-to-run variance, and outperforms prompt-only optimization. The discovered pipelines are objective-dependent: only under maximin does the researcher inject an explicit \emph{fairness mechanism} into synthesizer pipelines, a class of mechanism that is absent from its own objective-agnostic system prompt and from every efficiency-optimized pipeline. This supports an \emph{information-design} reading in which the researcher chooses what to reveal to the boundedly rational synthesizer as a function of the welfare objective. Code at \url{https://github.com/vicgalle/autoresearch-social-dilemmas}.
\end{abstract}

\section{Introduction}

Sequential Social Dilemmas (SSDs)~\cite{leibo2017} are the multi-agent analogue of the prisoner's dilemma in temporally rich Markov games: individually rational play leads to collectively suboptimal outcomes through pollution, over-harvesting, or open conflict. Standard multi-agent reinforcement learning (MARL) struggles in this regime due to credit assignment, non-stationarity, and large joint action spaces~\cite{busoniu2008}. A complementary approach, recently introduced by Gallego~\cite{gallego2026}, sidesteps these difficulties by replacing \emph{decentralized} parameter-space optimization with \emph{centralized} algorithm-space synthesis: a frozen LLM writes a Python policy function, evaluates it in self-play, and iteratively refines it from performance feedback. A single generation step can produce coordination logic (territory partitioning, role assignment, conditional cooperation) at a sample efficiency several orders of magnitude beyond what gradient-based MARL achieves on the same environments.

This shifts where the design problem lives, rather than removing it. The inner-loop pipeline that drives the synthesizer has many free parameters: which system prompt, which feedback variables, which helper functions, how many refinement steps. Each materially affects the resulting policies, and prior work tuned them by hand. A natural question follows: \emph{can an AI agent design the pipeline?}

We answer affirmatively with a two-level autoresearch framework. An \emph{outer-loop researcher agent} (Claude Opus 4.6, run as a coding agent) edits the source files of an \emph{inner-loop policy synthesizer} (another LLM), runs evaluations on held-out seeds, and keeps modifications that improve a fixed welfare objective $\Phi$. The outer agent operates on an ordinary git repository (reading code, writing diffs, running shell commands, etc) without task-specific scaffolding beyond a standard CLI and git, mirroring the autoresearch paradigm of Karpathy~\cite{karpathy2026} for single-GPU LLM pretraining. Although the inner-loop SSDs are gridworld benchmarks rather than physical systems, the outer-loop discovery process itself runs under conditions a deployed discovery agent faces: noisy multi-seed evaluations, stochastic code generation, an LLM-evaluation budget that bounds how often $\Phi$ can be queried, and a heterogeneous code repository the agent must navigate end-to-end on its own.

Our contributions are: i) a general two-level framework that delegates the design of an LLM synthesis pipeline to a coding agent operating on a real software repository (Section~\ref{sec:framework}); ii) the first instantiation of the autoresearch paradigm in a multi-agent decision-making domain, with experiments across two SSDs (Cleanup and Gathering), two policy LLMs, and two welfare objectives (utilitarian efficiency $U$ and Rawlsian maximin $\min_i R_i$) (Section~\ref{sec:experiments}); and iii) a mechanism-design interpretation supported by the qualitatively different pipelines the agent produces under different welfare objectives, including the autonomous insertion of explicit fairness mechanisms (usually \emph{time-based duty rotation}) into the researcher-authored synthesizer prompts and helpers, in every maximin run and no efficiency run.

\section{Background}

We build on the iterative LLM policy synthesis framework of Gallego~\cite{gallego2026}, which serves as the (frozen) inner loop of our two-level system. This section recalls the SSD formalism, the social outcome metrics, and the base synthesis loop.

\subsection{Sequential Social Dilemmas}

A Sequential Social Dilemma is a partially observable Markov game $\cG = \langle N, \mathcal{S}, \{\mathcal{A}_i\}_{i=1}^{N}, T, \{R_i\}_{i=1}^{N}, H \rangle$ with $N$ agents, state space $\mathcal{S}$ (the gridworld configuration), per-agent action spaces $\mathcal{A}_i$, transition function $T$, reward functions $R_i$, and episode horizon $H$~\cite{leibo2017}. Beyond the dilemma's matrix-game structure, SSDs add temporal richness: agents must learn \emph{when} and \emph{where} to cooperate, not just \emph{whether} to.

We study two canonical SSDs that capture complementary dilemma types.

\textbf{Cleanup}~\cite{hughes2018} is a \emph{public goods provision} game ($N{=}10$). The map has two regions: a river that accumulates waste, and an orchard where apples grow. Apples regrow only when river pollution is below threshold. Each agent can fire a cleaning beam (cost $-1$) that removes waste, collect apples ($+1$ each), or fire a tagging beam (cost $-1$, inflicting $-50$ on the target and removing it for $25$ steps). The dilemma: cleaning is costly but its benefits are public, so purely self-interested agents free-ride.

\textbf{Gathering}~\cite{leibo2017,perolat2017} is a \emph{common pool resource} game ($N{=}4$). Agents collect shared apples on a fixed respawn timer and may fire tagging beams to temporarily remove rivals. The dilemma: agents can coexist and share resources, or aggress to monopolize them; aggression wastes time and reduces total welfare.

Both games use $8$--$9$ discrete actions (movement, rotation, beam, stand, optionally clean) and episodes of $H{=}1000$ steps. The two dilemmas differ in cost structure: asymmetric provision (cleaners pay, all benefit) vs.\ symmetric restraint (every agent faces the same temptation), a distinction that drives our experimental findings.

\paragraph{Social metrics.} Following Perolat et al.~\cite{perolat2017}, let $R_i = \sum_{t=0}^{H-1} r_i^t$ denote agent $i$'s episode return. We evaluate four social outcomes:
\begin{align*}
U &= \tfrac{1}{H}\sum_{i=1}^{N} R_i & \text{(efficiency)} \\
E &= 1 - \frac{\sum_{i,j} |R_i - R_j|}{2N \sum_i R_i} & \text{(equality)} \\
S &= \tfrac{1}{N}\sum_{i=1}^{N} \bar t_i & \text{(sustainability)} \\
P &= \tfrac{1}{H}\sum_{t=0}^{H-1} \big|\{i : \textrm{active}_i^t\}\big| & \text{(peace)}
\end{align*}
where $\bar t_i$ is the mean timestep at which agent $i$ collects positive reward (higher $\Rightarrow$ resources preserved later) and $\textrm{active}_i^t$ indicates that agent $i$ is not tagged out at step $t$. We additionally consider the \emph{maximin} (Rawlsian) welfare criterion $\min_i R_i$, which optimizes the worst-off agent's return and serves as the second objective in our experiments.

\paragraph{A note on the dilemma's status under symmetric programmatic policies.}
We adopt the SSD environments of \cite{leibo2017,hughes2018,perolat2017} as benchmarks, but in our synthesis setup a single Python function $\pi$ controls all $N$ agents (\S\ref{sec:framework}). This reframes the strategic problem: the individual-rationality constraint that makes classical SSDs a \emph{dilemma} is replaced by a joint coordination/scheduling problem with the welfare objective $\Phi$ as the explicit target. Locally myopic per-agent code can still recreate dilemma-shaped behavior (and the baseline pipeline in fact does), but cooperation here is a joint-optimization outcome, not an equilibrium under individual rationality. We interpret the mechanisms the researcher discovers (duty rotation, role assignment) accordingly: they are coordination solutions in algorithm space that resemble the fairness mechanisms one would want a decentralized MARL system to converge to, not equilibria induced by self-interested agents.

\subsection{Iterative LLM Policy Synthesis}

Let $\Pi$ denote the space of \emph{code-based policies}: deterministic functions $\pi: \mathcal{S} \times [N] \to \mathcal{A}$ expressed as executable Python code. Each policy has access to the full environment state and a library of helpers (BFS pathfinding, beam targeting, coordinate transforms). This state access is a deliberate design choice: programmatic policies operate in algorithm space rather than the reactive observation-to-action space of neural policies, which lets a single LLM generation step encode rich coordination logic.

A frozen LLM $\cM$ acts as the \emph{policy synthesizer}. Given a system prompt $p$ describing the environment API and a feedback prompt $q_k$, it produces a new policy
\[
  \pi_{k+1} = \cM\!\left(p,\; q(\pi_k, \mathcal{F}_k)\right),
\]
where $\pi_k$ is the previous policy (its source code) and $\mathcal{F}_k$ is the evaluation feedback. All $N$ agents execute the same program $\pi_k$ in self-play. We stress that this is \emph{symmetric}, not \emph{behaviorally homogeneous}: since $\pi_k$ takes \texttt{agent\_id} as an argument, a single shared program can induce distinct per-agent behaviors (cleaner vs.\ gatherer assignment, time-rotated duty cycles, partitioned territories; see the synthesized policies in Appendix~\ref{app:policies}). What is shared is the source code; the sampled action distributions can differ across agents. Evaluation over a set of random seeds $S$ yields the mean per-agent return $\bar r_k$ and the social metrics vector $\bm_k = (U_k, E_k, S_k, P_k)$. Each generated policy passes an AST-based safety check (blocking \texttt{eval}, file I/O, network access) followed by a short smoke test; failures trigger regeneration (up to $R$ attempts) with the error message appended to the prompt.

\paragraph{Feedback.} We package the previous policy's code together with all available evaluation signals:
\begin{equation}
  \mathcal{F}_k = \big(\textrm{code}(\pi_k),\; \bar r_k,\; \bm_k,\; \mathbf{d}\big),
\end{equation}
where $\mathbf{d}$ contains natural-language definitions of each social metric. The LLM consumes $\mathcal{F}_k$ to revise and improve the policy. This is a starting point; the choice of feedback content is a single design decision within a much larger pipeline configuration space: our two-level framework (Section~\ref{sec:framework}) opens the full space to automated search.

\section{Two-Level Framework}
\label{sec:framework}
\begin{figure}[t]
\centering
\begin{minipage}[t]{0.45\linewidth}
\vspace{0pt}
\centering
\begin{tikzpicture}[
  >={Stealth[length=3.5pt]},
  node distance=0.4cm,
  mbox/.style={
    rectangle, rounded corners=3.5pt, draw=#1!85!black, fill=#1!15,
    line width=0.7pt, minimum height=0.55cm, minimum width=3.6cm,
    align=center, inner sep=2.5pt, font=\footnotesize},
  sbox/.style={
    rectangle, rounded corners=2pt, draw=ilcolor!65!black, fill=ilcolor!10,
    line width=0.5pt, minimum height=0.38cm,
    align=center, inner sep=1.5pt, font=\scriptsize},
  frzbox/.style={
    rectangle, rounded corners=2pt, draw=frzcolor!60!black, fill=frzcolor!8,
    dashed, line width=0.45pt, align=center, inner sep=2pt, font=\scriptsize},
  myarr/.style={->, line width=0.5pt, #1!60!black},
  myarr/.default=black,
  albl/.style={font=\scriptsize, text=black!50},
]
\node[mbox=rescolor] (R) {%
  \textbf{Researcher} $\cR$%
  \;\;{\scriptsize\textit{(Claude Opus 4.6)}}};
\node[mbox=pipcolor, below=0.42cm of R] (P) {%
  \textbf{Pipeline} $c_j$%
  \;\;{\scriptsize $p,\;\phi,\;\mathcal{H},\;\iota$}};
\node[sbox, below=0.8cm of P] (V) {Validate};
\node[sbox, left=0.12cm of V] (S) {Synthesize $\cM$};
\node[sbox, right=0.12cm of V] (Ev) {Evaluate in $\cG$};
\node[sbox, minimum width=0.9cm, below=0.28cm of V] (FB) {Feedback $\mathcal{F}_k$};
\begin{scope}[on background layer]
\node[draw=ilcolor!35, fill=ilcolor!3, rounded corners=5pt, line width=0.55pt,
      inner xsep=5pt, inner ysep=4pt,
      fit=(S)(V)(Ev)(FB),
      label={[font=\scriptsize\itshape, text=ilcolor!50!black]%
             above:{Inner Loop \;($K$ iterations)}}
      ] (IL) {};
\end{scope}
\node[mbox=metcolor, below=0.42cm of IL] (M) {%
  Metrics:\; $J_j$,\; $\mathbf{m}_j$,\; $\Delta_j$};
\node[mbox=deccolor, below=0.38cm of M] (D) {%
  $J_j > J^* + \tau$\,?\;\;%
  {\scriptsize\textcolor{metcolor!80!black}{\textbf{keep}}%
   \;$\mid$\;%
   \textcolor{deccolor!80!black}{\textbf{discard}}}};
\draw[myarr] (R) -- node[albl, right] {modify} (P);
\draw[myarr] (P) -- (IL);
\draw[myarr] (IL) -- (M);
\draw[myarr] (M) -- (D);
\draw[myarr=ilcolor] (S) -- (V);
\draw[myarr=ilcolor] (V) -- (Ev);
\draw[myarr=ilcolor, rounded corners=2pt] (Ev.south) |- (FB);
\draw[myarr=ilcolor, rounded corners=2pt] (FB) -| (S.south);
\draw[myarr=rescolor, rounded corners=5pt, line width=0.55pt]
  (D.west) -- ++(-1.1,0) |-
  node[albl, left, pos=0.25] {history} (R.west);

\begin{scope}[on background layer]
\draw[rescolor!30, dashed, rounded corners=7pt, line width=0.65pt]
  ([shift={(-1.35cm,0.3cm)}]R.north west) rectangle
  ([shift={(0.25cm,-0.35cm)}]D.south east);
\end{scope}
\node[font=\tiny\itshape, text=rescolor!45!black, anchor=south east]
  at ([shift={(0.2cm,-0.3cm)}]D.south east) {$j = 1,\ldots,J$};
\end{tikzpicture}
\captionof{figure}{Two-level automated research framework (Algorithm~\ref{alg:outer}).}
\label{fig:framework}
\end{minipage}\hfill
\begin{minipage}[t]{0.52\linewidth}
\vspace{0pt}
\begin{algorithm}[H]
\caption{Two-Level Automated Research}
\label{alg:outer}
\begin{algorithmic}[1]
\footnotesize
\Require Game $\cG$, policy synthesizer $\cM$, researcher $\cR$, system prompt $p_{\cR}$, initial config $c_0$, outer iterations $J_{\max}$, welfare objective $\Phi$, held-out seeds $S_{\textrm{ho}}$, keep threshold $\tau \geq 0$
\Ensure Best configuration $c^*$, best policy $\pi^*$
\State $\pi^*_0 \leftarrow \textsc{InnerLoop}(\cM, \cG, c_0)$
\State $J_0 \leftarrow \Phi(\textsc{Eval}(\pi^*_0;\, \cG,\, S_{\textrm{ho}}))$
\State $c^* \leftarrow c_0,\; J^* \leftarrow J_0$ \hfill {\scriptsize\textit{// running best}}
\State $\textit{history} \leftarrow \{(c_0, J_0, \bm_0, \varnothing)\}$
\For{$j = 1, \dots, J_{\max}$}
  \State $c_j \leftarrow \cR\!\left(p_{\cR},\, \textsc{code}(c^*),\, \textit{history}\right)$
  \State \textbf{if} $\neg\,\textsc{ValidateConfig}(c_j)$ \textbf{then} retry $\leq R$
  \State $\pi^*_j \leftarrow \textsc{InnerLoop}(\cM, \cG, c_j)$
  \State $J_j \leftarrow \Phi(\textsc{Eval}(\pi^*_j;\, \cG,\, S_{\textrm{ho}}))$
  \State $\Delta_j \leftarrow \textsc{Diff}(c^*, c_j)$ \hfill {\scriptsize\textit{// code diff}}
  \State \textbf{if} $J_j > J^* + \tau$ \textbf{then keep:} $c^* \leftarrow c_j,\; J^* \leftarrow J_j$;\; \textbf{else discard} \hfill {\scriptsize\textit{// $\tau=0$ in our runs}}
  \State $\textit{history} \leftarrow \textit{history} \cup \{(c_j, J_j, \bm_j, \Delta_j)\}$
\EndFor
\State \Return $c^*,\, \pi^*_{c^*}$
\end{algorithmic}
\end{algorithm}
\end{minipage}
\end{figure}
We introduce a two-level system where a \emph{researcher agent} $\cR$ autonomously discovers configurations that optimize the output of an inner-loop system. While we instantiate this for multi-agent policy synthesis, the architecture is general: any pipeline where an LLM generates artifacts, evaluates them, and iterates can serve as the inner loop. The fundamental insight is that the entire inner-loop codebase is a designable artifact that a code-based agent can search over. Figure~\ref{fig:framework} illustrates the architecture.

\subsection{Configuration Space}

Let $\cC$ denote the space of \emph{pipeline configurations}. Each configuration $c \in \cC$ specifies the full inner-loop setup:
\begin{equation}
  c = (p,\, \phi,\, \mathcal{H},\, \iota)
\end{equation}
where $p$ is the system prompt, $\phi$ is the feedback construction function (which metrics and diagnostics to include, how to frame it, whether to inject adaptive hints and thresholds, etc), $\mathcal{H}$ is the helper function library (auxiliary functions for pathfinding, getting aggregates of useful environemnt quantities, etc.), and $\iota$ specifies the iteration logic (number of inner iterations $K$, sampling strategy). Table~\ref{tab:config} provides concrete examples. The validation pipeline is part of the frozen inner-loop infrastructure rather than a configurable component, the researcher cannot modify it in our experiments to prevent reward hacking.

\begin{table}[t]
\caption{Configuration components modifiable by the researcher, with concrete edits that $\cR$ made in our runs. The environment simulator, ground-truth evaluation, and policy LLM weights are frozen.}
\label{tab:config}
\centering
\footnotesize
\begin{tabular}{@{}llp{0.60\linewidth}@{}}
\toprule
 & Scope & Concrete edits by $\cR$ (see appendix) \\
\midrule
$p$ & System prompt & Multi-section strategic briefing with explicit duty-rotation template \texttt{(agent\_id + step//T) \% n} (App.~\ref{app:art-prompts}, Listing~\ref{lst:prompt-min}) \\
$\phi$ & Feedback fn & Thresholded diagnostics: ``FAIRNESS ALERT'' fires when $\min_i R_i {<} 0$; ``DO NOT REGRESS'' guard fires when $U {\geq} 2.5$ (App.~\ref{app:art-feedback}, Listing~\ref{lst:feedback-adapt}) \\
$\mathcal{H}$ & Helper library & BFS-Voronoi territories with respawn-timer awareness; band-based apple zoning by \texttt{agent\_id} (App.~\ref{app:art-helpers}, Listings~\ref{lst:helper-vor},~\ref{lst:helper-zone}) \\
$\iota$ & Iteration logic & Per-condition $K{\in}\{2,3\}$; $|S|$ walked $5{\to}8{\to}12$ for variance control; thinking budget $10$--$32$k (App.~\ref{app:art-config}, Table~\ref{tab:iota}) \\
\bottomrule
\end{tabular}
\end{table}

The hand-designed feedback of~\cite{gallego2026} corresponds to a single fixed instantiation of $\phi$. Our framework opens the full configuration space to automated search.

\subsection{Inner Loop (Policy Synthesis)}

Given a configuration $c$, the inner loop executes $K$ iterations of LLM policy synthesis:
\begin{equation}
  \pi^*_c = \textsc{InnerLoop}(\cM, \cG, c)
\end{equation}
Each iteration $k$ proceeds in four stages, following~\cite{gallego2026}:
\begin{enumerate}
  \item \textbf{Synthesize.} The policy LLM $\cM$ receives the system prompt $p$, the previous policy's source code $\pi_{k-1}$, and feedback $\mathcal{F}_{k-1}$ constructed by $\phi$. It generates a new Python policy function $\pi_k$ that has access to full environment state and the helper library $\mathcal{H}$.
  \item \textbf{Validate.} The generated code undergoes AST-based safety checks (blocking dangerous operations such as file I/O and network access) followed by a short smoke test. Failures trigger re-generation (up to $R$ retries), with the error message appended to the prompt.
  \item \textbf{Evaluate.} All $N$ agents execute the same policy $\pi_k$ in self-play over $|S|$ random seeds (note the policy is conditional on \texttt{agent\_id}). The evaluation yields the mean per-agent reward $\bar{r}_k$ and the social metrics vector $\bm_k = (U_k, E_k, S_k, P_k)$.
  \item \textbf{Feedback.} The feedback function $\phi$ constructs the prompt for the next iteration from $(\pi_k, \bar{r}_k, \bm_k)$, packaging the previous policy's code together with the scalar reward, the social metrics vector, their natural-language definitions, and any adaptive diagnostics that $\phi$ injects.
\end{enumerate}

The inner loop output is scored on held-out seeds via the configuration-level map
\begin{equation}
  J(c) = \Phi\!\left(\textsc{Eval}(\pi^*_c;\; \cG,\; S_{\text{held-out}})\right),
  \label{eq:J}
\end{equation}
where $\textsc{Eval}$ returns the per-agent returns of $\pi^*_c$ in $\cG$ averaged over the held-out seeds, and $\Phi$ is a fixed welfare functional that aggregates those returns into a scalar. We consider two alternative welfare functionals:
\begin{align}
  \Phi_U &= U = \frac{1}{H}\sum_{i=1}^{N} R_i & \text{(utilitarian efficiency)} \label{eq:eff}\\
  \Phi_{\min} &= \min_i R_i & \text{(Rawlsian maximin)} \label{eq:rawls}
\end{align}
$\Phi_U$ rewards collective throughput and is indifferent to how reward is distributed across agents, whereas $\Phi_{\min}$ instead optimizes for the worst-off agent, pressuring the researcher toward configurations that distribute the cost of cooperation. The researcher's goal is to maximize $J(c)$ for a chosen $\Phi$; we use $\Phi$ to denote the welfare objective throughout, and $J$ for the per-configuration scalar score returned by held-out evaluation.

\subsection{Outer Loop (Automated Research)}

The researcher agent $\cR$ iteratively modifies the pipeline configuration. Following the autoresearch paradigm~\cite{karpathy2026}, $\cR$ operates on the inner-loop codebase as a modifiable artifact, proposing changes, observing outcomes, and refining. The procedure is formalized in Algorithm~\ref{alg:outer}.

At each outer iteration $j$, the researcher $\cR$ receives: i) the \textbf{full source code} of the current running-best configuration $c^*$ (prompts, feedback construction, helpers, iteration logic); ii) the \textbf{experiment history}: for each prior iteration, the code diff $\Delta_i$, ground-truth score $J_i$, social metrics vector $\bm_i$, and whether the iteration was kept or discarded; iii) the \textbf{environment source code} (read-only), enabling the researcher to reason about game mechanics. Discarded iterations are reverted on disk (\texttt{git checkout -- pipeline/}) so that the next proposal $c_{j+1}$ is constructed on top of $c^*$, not on top of $c_j$.
The researcher proposes a new configuration $c_j$ by generating code modifications. Concretely, $\cR$ is a coding agent (Claude Code CLI) that operates on a real software repository: it reads and edits Python source files, runs shell commands, inspects evaluation outputs, and commits changes to a dedicated git branch, following the same workflow a human researcher would follow.

\subsection{Connection to Automated Mechanism Design}
\label{sec:mechdesign}

The two-level structure admits a mechanism design interpretation. The researcher $\cR$ acts as a \emph{mechanism designer}: it controls the information structure (what metrics to reveal, how to frame them), the action space (what helper functions are available), and the incentive structure (how feedback is presented) under which the policy synthesizer $\cM$ operates. The synthesizer acts as the \emph{agent} within the designed mechanism.

This connects to the automated mechanism design literature~\cite{conitzer2002}, where a principal designs rules to induce desired behavior from self-interested agents. In our setting: (i) the \textbf{principal} is the researcher $\cR$, optimizing the ground-truth score $J$ induced by the welfare objective $\Phi$; (ii) the \textbf{agent} is the synthesizer $\cM$, optimizing per-agent reward as instructed; (iii) the \textbf{mechanism} is the configuration $c$: prompts, feedback, helpers, iteration logic; (iv) the \textbf{outcome} is the social welfare of the resulting multi-agent policy $\pi^*_c$.

A crucial distinction from classical mechanism design: $\cM$ follows instructions but has \emph{bounded rationality} in the sense that its ability to synthesize effective policies depends on the information and tools provided. The researcher's task is thus closer to \emph{information design}~\cite{kamenica2011}: choosing what to reveal to help $\cM$ navigate the cooperation--defection tension. Our experiments (Section~\ref{sec:experiments}) show that the researcher designs qualitatively different information structures depending on the welfare objective $\Phi$, supporting this interpretation empirically.

\section{Experiments}
\label{sec:experiments}


We conduct 12 autonomous researcher runs across a factorial design. The researcher agent $\cR$ is Claude Opus 4.6, invoked via the Claude Code CLI as a coding agent. Each run operates on a dedicated git branch of a real Python codebase: $\cR$ edits source files in \texttt{pipeline/}, executes evaluation scripts, reads metric outputs, and iterates, without human intervention.

\paragraph{Design.}
For Cleanup: $2$ policy LLMs $\times$ $2$ objectives $\times$ $2$ replications $=$ 8 runs. For Gathering: $2$ policy LLMs $\times$ $1$ objective $\times$ $2$ replications $=$ 4 runs. Maximin runs are unnecessary for Gathering because efficiency optimization alone achieves close to perfect equality.

\paragraph{Models.}
Policy synthesizer $\cM$: Gemini 3.1 Pro (Google) or Claude Sonnet 4.6 (Anthropic), to recent state-of-the-art LLMs. Both use extended thinking. We additionally test with Gemma 4 26B-A4B-IT (Google), a smaller open-weight model, to probe the framework's behavior when the policy synthesizer has substantially lower capability (Appendix~\ref{app:gemma}).

\paragraph{Baselines.}
The hand-designed feedback configuration from prior work~\cite{gallego2026} serves as the initial pipeline $c_0$ for all runs. On Cleanup ($N{=}10$), this baseline achieves $U{=}1.93$/$2.70$ (mean/max) with Gemini and $U{=}0.86$/$1.56$ with Sonnet. We additionally compare against GEPA~\cite{agrawal2026}, an automated prompt optimization method that iteratively refines the system prompt via LLM reflection. GEPA optimizes only the system prompt $p$, whereas our researcher modifies the full pipeline $(p, \phi, \mathcal{H}, \iota)$. We give GEPA a matched compute budget: the same number of optimization steps as outer iterations used by our method, so all in all both methods use a comparable number of environment evaluations.

\subsection{Results}

Table~\ref{tab:results} presents the main results, comparing our autoresearch framework against the hand-designed baseline of~\cite{gallego2026} and GEPA~\cite{agrawal2026}.

\begin{table}[t]
\caption{Results. Mean\,{\scriptsize/\,max} across the runs per condition. \textbf{Bold} marks the best mean per metric within each Policy LLM $\times$ Game sub-block. Baseline: unmodified, hand-designed pipeline from~\cite{gallego2026}. GEPA: automated prompt optimization~\cite{agrawal2026} with matched compute budget.}
\label{tab:results}
\centering
\footnotesize
\begin{tabular}{@{}llccc@{}}
\toprule
Policy LLM & Target & $U$ & $E$ & $\min_i R_i$ \\
\midrule
\multicolumn{5}{@{}l}{\emph{Cleanup --- Baseline}} \\
Gemini 3.1 Pro & --- & 1.93{\scriptsize/2.70} & 0.17{\scriptsize/0.62} & $-$159{\scriptsize/$-$84} \\
Sonnet 4.6 & --- & 0.86{\scriptsize/1.56} & $-$0.02{\scriptsize/0.25} & $-$151{\scriptsize/$-$59} \\
\addlinespace
\multicolumn{5}{@{}l}{\emph{Cleanup --- GEPA~\cite{agrawal2026}}} \\
\multirow{2}{*}{Gemini 3.1 Pro}
  & $\Phi_U$ & 1.34{\scriptsize/1.37} & $-$0.10{\scriptsize/$-$0.05} & $-$149{\scriptsize/$-$126} \\ 
  & $\Phi_{\min}$ & 1.76{\scriptsize/2.76} & 0.90{\scriptsize/0.96} & 143{\scriptsize/245} \\
\addlinespace
\multirow{2}{*}{Sonnet 4.6}
  & $\Phi_U$ & 1.04{\scriptsize/1.20} & $-$0.59{\scriptsize/$-$0.21} & $-$164{\scriptsize/$-$126} \\
  & $\Phi_{\min}$ & $-$0.63{\scriptsize/1.60} & 0.77{\scriptsize/1.00} & $-$147{\scriptsize/6} \\
\addlinespace
\multicolumn{5}{@{}l}{\emph{Cleanup --- Two-level Autoresearch}} \\
\multirow{2}{*}{Gemini 3.1 Pro}
  & $\Phi_U$ & \textbf{3.20}{\scriptsize/3.25} & 0.55{\scriptsize/0.61} & $-$211{\scriptsize/$-$182} \\
  & $\Phi_{\min}$ & 3.16{\scriptsize/3.19} & \textbf{0.98}{\scriptsize/0.98} & \textbf{290}{\scriptsize/296} \\
\addlinespace
\multirow{2}{*}{Sonnet 4.6}
  & $\Phi_U$ & \textbf{3.12}{\scriptsize/3.14} & 0.66{\scriptsize/0.70} & $-$196{\scriptsize/$-$146} \\
  & $\Phi_{\min}$ & 2.57{\scriptsize/2.93} & \textbf{0.91}{\scriptsize/0.97} & \textbf{179}{\scriptsize/200} \\
\midrule
\multicolumn{5}{@{}l}{\emph{Gathering --- Baseline}} \\
Gemini 3.1 Pro & --- & 2.04{\scriptsize/2.42} & 0.90{\scriptsize/0.98} & 412{\scriptsize/571} \\
Sonnet 4.6 & --- & 0.03{\scriptsize/0.03} & 0.54{\scriptsize/0.54} & 0{\scriptsize/0} \\
\addlinespace
\multicolumn{5}{@{}l}{\emph{Gathering --- GEPA~\cite{agrawal2026}}} \\
Gemini 3.1 Pro & $\Phi_U$ & 2.08{\scriptsize/2.35} & 0.94{\scriptsize/0.96} & 436{\scriptsize/518} \\  
Sonnet 4.6 & $\Phi_U$ & 1.20{\scriptsize/1.23} & 0.63{\scriptsize/0.71} & 86{\scriptsize/123} \\
\addlinespace
\multicolumn{5}{@{}l}{\emph{Gathering --- Two-level Autoresearch}} \\
Gemini 3.1 Pro & $\Phi_U$ & \textbf{2.49}{\scriptsize/2.51} & \textbf{0.98}{\scriptsize/0.98} & \textbf{582}{\scriptsize/582} \\
Sonnet 4.6 & $\Phi_U$ & \textbf{2.52}{\scriptsize/2.52} & \textbf{0.96}{\scriptsize/0.98} & \textbf{576}{\scriptsize/597} \\
\bottomrule
\end{tabular}
\end{table}

\begin{figure}[t]
\centering
\includegraphics[width=0.79\textwidth]{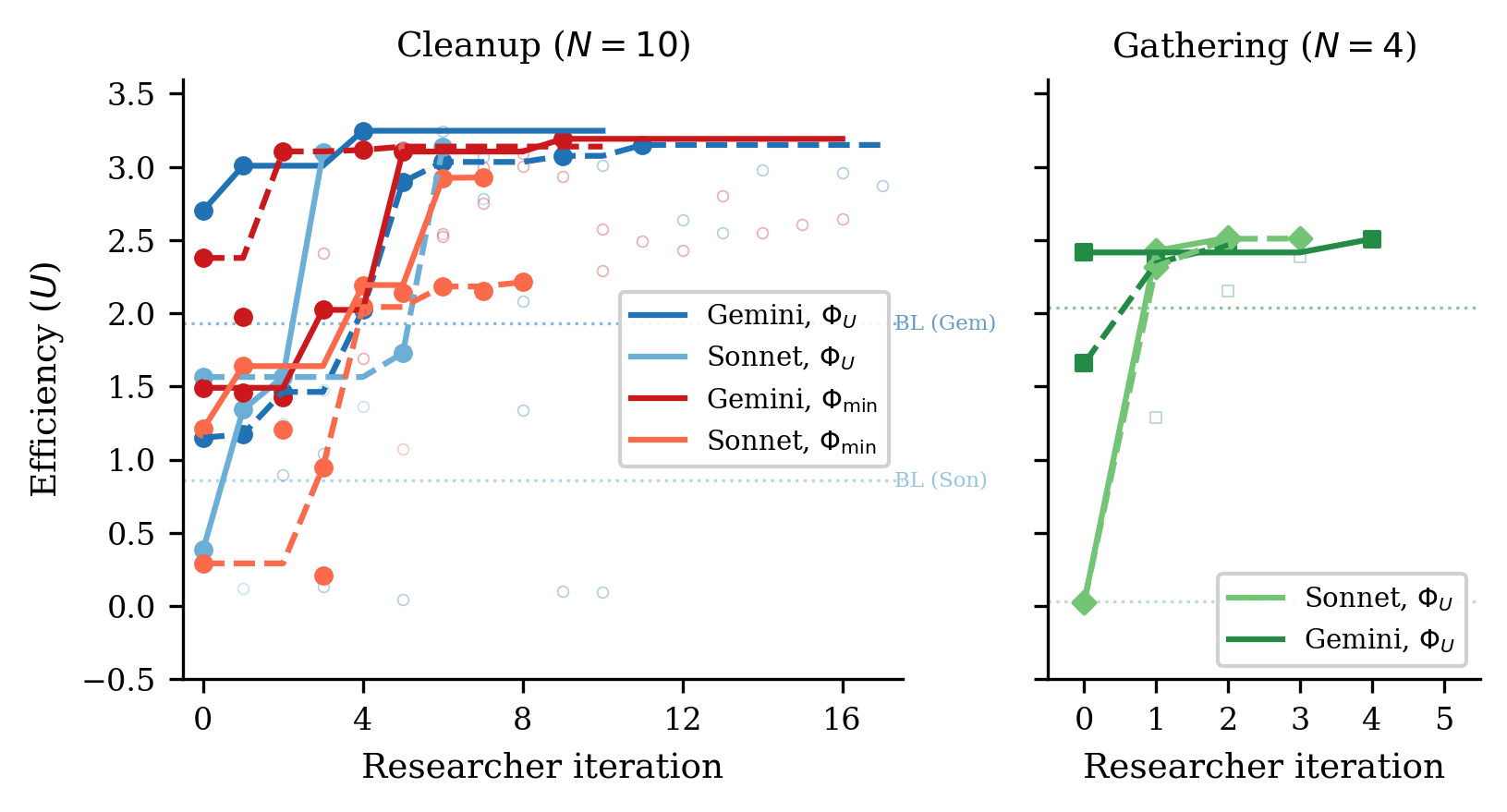}
\caption{Efficiency ($U$) across researcher iterations for all 12 runs. \emph{Left}: Cleanup with 8 runs across 2 LLMs $\times$ 2 objectives. Solid lines connect \emph{kept} iterations (those for which $J_j > J^* + \tau$ strictly exceeded the running-best score; $\tau{=}0$ in our runs); open circles mark \emph{discarded} iterations ($J_j \le J^*$), which are reverted on disk before the next proposal. Dashed horizontal line: hand-designed baseline from~\cite{gallego2026}. All runs converge to $U \approx 3.1$--$3.2$ despite diverse starting points. \emph{Right}: Gathering with 4 runs (2 per LLM). All converge to $U \approx 2.5$ within 2--4 iterations.}
\label{fig:efficiency}
\end{figure}

\paragraph{Finding 1: The researcher reliably improves over hand-designed baselines and outperforms prompt-only optimization.}
Every run improves substantially regardless of starting point (Figure~\ref{fig:efficiency}). On Cleanup, autoresearch lifts both LLMs to $U \approx 3.1$--$3.2$ from baselines of $1.93$ (Gemini) and $0.86$ (Sonnet), nearly closing the gap between them; on Gathering, all four runs converge to $U \in [2.47, 2.52]$ from baselines spanning $0.03$--$2.42$. Run-to-run spread is tight (gaps within $0.05$ on Cleanup-$\Phi_U$), suggesting the researcher reliably finds the performance ceiling of each policy LLM via the pipeline modifications it discovers (helpers, prompts, and feedback; see appendices ~\ref{app:art-helpers}--\ref{app:art-feedback} for a selection of them). At matched environment queries, autoresearch beats GEPA by $2$--$3{\times}$ on Cleanup (both $\Phi_U$ and $\Phi_{\min}$) and $20\%$ on Gathering, with the gap widening for the weaker policy LLM: GEPA-Sonnet under $\Phi_{\min}$ can collapse to a pathological ``everyone cleans, nobody eats'' regime ($U{=}{-}2.87$, $E{=}1.00$), while autoresearch-Sonnet reaches $\min_i R_i \approx 200$ reliably. Modifying the full pipeline, not just the prompt, is what closes these gaps.

\paragraph{Finding 2: No efficiency--fairness tradeoff in Cleanup (Gemini).}
Maximin-optimized Gemini pipelines sacrifice only $1\%$ efficiency ($U$: $3.16$ vs.\ $3.20$) while achieving near-perfect equality ($E$: $0.98$ vs.\ $0.55$) and transforming maximin from deeply negative baselines ($-99$ to $-84$) to $\min_i R_i = 290$ (Figure~\ref{fig:maximin}). The researcher discovers \emph{fair duty rotation} (Listing~\ref{lst:min-gemini}; primed by the prompt rewrite of Listing~\ref{lst:prompt-min})---time-based cycling using \texttt{agent\_id} and \texttt{env.\_step\_count}---simultaneously improving worst-off welfare \emph{and} collective output. Because cleaning is a public good, distributing the cleaning cost fairly ensures enough cleaners to sustain apple production.

For Sonnet, there is a moderate tradeoff: efficiency drops from $3.12$ to $2.57$ under maximin optimization. The gap reflects Sonnet's harder time implementing complex coordination mechanisms (role rotation, zone assignment) from strategic hints alone.

\paragraph{Finding 3: Game structure determines whether fairness requires explicit optimization.}
In Cleanup, where cleaning costs are borne asymmetrically (cleaners pay $-1$, free-riders collect apples), baseline equality ranges from $E{=}0.04$ to $0.62$, and maximin optimization is required to reach $E > 0.9$. In Gathering, where all agents face a symmetric landscape, efficiency optimization alone achieves $E > 0.94$ across all 4 runs: no separate maximin runs are needed. This generalizes: in \emph{provision} dilemmas with asymmetric costs, fairness requires designed mechanisms (role rotation, duty sharing); in \emph{restraint} dilemmas with symmetric costs, fairness emerges as a free byproduct of efficient coordination. The researcher independently discovers this, it creates role differentiation pipelines only for Cleanup, and pure spatial-coordination pipelines for Gathering.

\paragraph{Finding 4: Convergent discovery of qualitatively different strategies per objective.}
Despite fully independent runs, the researcher converges on the same core strategies within each condition (Table~\ref{tab:strategies}, Appendix~\ref{app:additional}). In Cleanup, waste-counting helpers and spatial zone partitioning appear across all runs. The qualitative dividing line is the presence of an explicit \emph{fairness mechanism}, which appears in 4/4 maximin runs but 0/4 efficiency runs. In 3/4 maximin runs (both Gemini runs and one Sonnet run, Listings~\ref{lst:min-gemini},~\ref{lst:min-sonnet}) the researcher writes \emph{time-based role rotation} into the synthesizer prompt; in the remaining maximin run the researcher writes a structurally distinct ``collective threshold'' mechanism in which all agents synchronously switch between cleaning and collecting based on \texttt{waste\_fraction(env)} (achieving comparable maximin without an agent-index phase). Under efficiency optimization, the researcher instead writes \emph{static} role assignment (some agents always clean), producing high collective output at the cost of equality (Listing~\ref{lst:eff-gemini}). In Gathering, the researcher discovers BFS-Voronoi territory partitioning and respawn-timer awareness (Listings~\ref{lst:helper-vor},~\ref{lst:gather-gemini}), with no role differentiation: optimization is purely spatial (who collects which apples) and temporal (respawn-aware positioning).

The convergence here is at the level of \emph{which artifacts $\cR$ injects} into $p$, $\phi$, $\mathcal{H}$. Once the researcher-authored prompt contains a rotation template (e.g., Listing~\ref{lst:prompt-min}), the downstream synthesizer's implementation is unsurprising. The non-trivial claim is that $\cR$ writes such a template only under $\Phi_{\min}$, never under $\Phi_U$, despite the researcher system prompt $p_\cR$ being identical across objectives and containing no rotation formula, no objective-conditional guidance, and no link between any strategy class and either welfare criterion (Listing~\ref{lst:r-prompt}, Appendix~\ref{app:art-r-prompt}).

\paragraph{Common failure modes.}
Three named patterns drive most discarded iterations: (1)~\emph{over-prescription} (too many strategic hints confuse $\cM$), (2)~\emph{iteration regression} ($K{\in}\{4,5\}$ over-refines a working policy), and (3)~\emph{feedback overload} (verbose per-agent diagnostics cause over-correction). Counts per mode and the residual ``pure $J$-regression'' category are tabulated in Appendix~\ref{app:discards}.

\paragraph{Spec-gaming on the modifiable surface.}
The tuple $c=(p,\phi,\mathcal{H},\iota)$ is unconstrained, so $\cR$ could in principle game the held-out $J$ by exposing simulator internals, hard-coding actions, or overfitting hyperparameters to the eval seed set. We inspected the final pipelines from all 12 runs and found no such patterns: $\mathcal{H}$ edits are spatial heuristics and state queries, $\phi$ edits are thresholded interventions on the optimized metric itself (not on seed-specific values), and $\cR$ walks $|S|$ \emph{upward} ($5{\to}8{\to}12$) when chasing maximin (reducing seed-noise, not exploiting it). The full inspection, including the closest observed near-miss (over-prescription, which manifests as a regression on $J$ rather than an undeserved gain), is reported in Appendix~\ref{app:specgaming}.

\section{Related Work}

\paragraph{Sequential social dilemmas.}
SSDs were introduced by Leibo et al.~\cite{leibo2017} (Gathering) and extended to public-goods settings by Hughes et al.~\cite{hughes2018} (Cleanup); Perolat et al.~\cite{perolat2017} formalized the social outcome metrics $(U, E, S, P)$ used here. Subsequent work studies inequity aversion, intrinsic motivation, and reputational mechanisms for promoting cooperation in MARL agents. We complement this line by automating the search for cooperative \emph{programs} rather than evolving neural policies, and by treating the welfare objective $\Phi$ as a designable parameter that the outer agent optimizes for, obtaining qualitatively different cooperative behaviors (static role assignment vs.\ duty rotation) under different objectives without changing the environment.

\paragraph{LLMs for policy and program synthesis.}
FunSearch~\cite{romera2024} evolves programs for combinatorial discovery; Eureka~\cite{ma2024} synthesizes reward functions from environment source code; Voyager~\cite{wang2024} and Code as Policies~\cite{liang2023} generate executable skill code for single-agent embodied control; ReEvo~\cite{ye2024} evolves heuristics with reflective feedback. These works target single-agent settings or non-strategic optimization. Gallego~\cite{gallego2026}, on which our inner loop is based, extended LLM program synthesis to the multi-agent SSD setting, where one program must coordinate $N$ self-play copies. Our contribution is one level above: rather than tuning the synthesizer for one task, we let an outer agent rewrite the synthesis pipeline itself.

\paragraph{LLM reflection and prompt optimization.}
Reflexion~\cite{shinn2023}, Self-Refine~\cite{madaan2023}, OPRO~\cite{yang2024}, and GEPA~\cite{agrawal2026} demonstrate that structured verbal feedback loops improve LLM outputs; ERL~\cite{shi2026} internalizes such reflection via self-distillation. These methods optimize the prompt or the model's reasoning trajectory in isolation. Our framework instead modifies the entire surrounding pipeline (system prompt, feedback construction, helper library, iteration logic) making prompt optimization one component of a strictly larger search space. 

\paragraph{Automated AI research.}
A small but growing body of work delegates the design of ML pipelines to LLM coding agents. Karpathy's autoresearch~\cite{karpathy2026} runs a coding agent that modifies \texttt{train.py} for nanoGPT pretraining and is rewarded for validation loss. Our system shares this autoresearch architecture (coding agent + frozen evaluation harness + diff-based history) but targets a different inner loop (multi-agent policy synthesis instead of single-model pretraining) and a different objective (multi-agent social welfare instead of validation bits-per-byte). To our knowledge this is the first instantiation of the autoresearch paradigm in a multi-agent decision-making domain.

\paragraph{Automated mechanism and information design.}
Classical automated mechanism design~\cite{conitzer2002} computes optimal allocation rules for self-interested agents under explicit incentive constraints, while information design~\cite{kamenica2011} studies how a principal commits to a signaling policy that shapes a receiver's behavior. Our outer agent solves a related but distinct problem: $\cM$ is not strategically deceptive (it follows instructions) but is \emph{boundedly rational}, so the researcher must decide what information, helpers, and structure to expose so that $\cM$ writes policies achieving the principal's welfare goal. Section~\ref{sec:experiments} shows that the pipelines $\cR$ produces are genuinely a function of the welfare objective, supporting this information-design framing empirically.

\section{Discussion and Conclusion}

We have presented a two-level framework where a coding agent autonomously discovers pipeline configurations that improve the output of an inner-loop LLM system. Applied to multi-agent policy synthesis in social dilemmas, the researcher agent reliably exceeds hand-designed baselines across multiple independent runs, and converges on qualitatively similar strategies within each game--objective condition. The framework requires no task-specific scaffolding beyond a standard CLI and git: the agent operates on a standard software repository using the same tools (file editing, shell commands, git) available to a human researcher. The inner-loop validation pipeline, helper-library skeleton, and orchestrator API are themselves deliberately scoped artifacts (see App.~\ref{sec:limitations}).

\paragraph{Mechanism design in action.}
Our results empirically support the mechanism design interpretation of Section~\ref{sec:mechdesign}. The researcher acts as an information designer: under $\Phi_U$, it reveals efficiency-oriented information (waste counts, zone assignments) that guides $\cM$ toward productive but unequal role allocation (Listing~\ref{lst:eff-gemini}). Under $\Phi_{\min}$, it additionally reveals fairness-oriented structure such as rotation schedules and equity feedback (Listings~\ref{lst:prompt-min},~\ref{lst:feedback-adapt}), guiding $\cM$ toward egalitarian coordination.

\paragraph{Human oversight.}
Our system is fully autonomous by design, but its architecture offers natural affordances for human-in-the-loop oversight: $\cR$ operates on a standard git repository, fully-auditable.
More broadly, the separation between the researcher $\cR$ and the evaluation $\Phi$ creates a \emph{delegation boundary}: the human defines \emph{what} to optimize (the welfare objective), while the agent decides \emph{how}. 

\paragraph{Future work.}
First, we are interested in applying the two-level framework to other LLM-driven pipelines (code optimization, scientific experiment design, infrastructure tuning), where the same architecture applies with a different inner loop and objective $\Phi$; next, adversarial objectives can be intriguing, to test whether the researcher discovers exploitative pipeline configurations, exposing reward hacking risks; and asymmetric programs: extend to settings where agents run \emph{different} source code (the present setup is symmetric in code but already supports heterogeneous behavior via \texttt{agent\_id}).

\bibliographystyle{unsrt}
\bibliography{references}


\appendix

\section{Limitations}
\label{sec:limitations}

We flag three limitations of the present study, in roughly decreasing order of how much they bound the claims.

\begin{enumerate}[leftmargin=*]
  \item \emph{Single researcher LLM.} All main-experiment runs (and the Gemma appendix) use the same researcher $\cR$ (Claude Opus 4.6 via the Claude Code CLI). A researcher ablation across $\cR$-LLMs, in which the same fixed system prompt $p_\cR$ (Appendix~\ref{app:art-r-prompt}) is run with several frontier coding agents, is the single most important follow-up.
  \item \emph{Inner-loop infrastructure is itself a designed artifact.} The line ``$\cR$ uses no task-specific scaffolding'' refers to the researcher's tooling (CLI, git, file edits) and to the fact that the configuration $c{=}(p,\phi,\mathcal{H},\iota)$ starts from a deliberately weak baseline. It does \emph{not} mean the surrounding system is unscoped: the inner-loop validation pipeline (AST safety checks, sandboxed execution, multi-seed evaluation harness), the helper-library skeleton (\texttt{pipeline/helpers.py} starts non-empty), and the orchestrator API (\texttt{run\_inner\_loop.py}) are themselves engineered artifacts that bound what $\cR$ can and cannot do. Generalization of the framework to settings without an equivalent harness is plausible but unevaluated.
  \item \emph{Gridworld scope.} The inner-loop SSDs are 2D gridworlds with fully-observed integer states, discrete action spaces, and short ($H{=}1000$-step) episodes. The specific mechanisms the researcher discovers (BFS-Voronoi partitioning, time-rotated agent-id phase counters, waste-fraction threshold rules) exploit this structure and are gridworld-shaped; their transfer to higher-dimensional, continuous, or partially-observable multi-agent settings is an open question. The outer-loop \emph{operating conditions} (noisy multi-seed evaluation, code-level edits to a real repository, bounded $J$-query budget) are realistic, but the conclusions about which strategies emerge are bounded by the inner-loop benchmark.
\end{enumerate}

\section{Additional Results}
\label{app:additional}

\begin{table}[h]
\caption{Strategies discovered by the researcher in Cleanup across 4 efficiency-optimized and 4 maximin-optimized independent runs. The grouped row records any explicit \emph{fairness mechanism}; its two sub-rows decompose this into the dominant variant (time-based role rotation) and the alternative (synchronized whole-population clean/collect switching) found in one Sonnet maximin run.}
\label{tab:strategies}
\centering
\small
\begin{tabular}{@{}lcc@{}}
\toprule
Strategy & $\Phi_U$ (4 runs) & $\Phi_{\min}$ (4 runs) \\
\midrule
Waste-counting helpers & 4/4 & 4/4 \\
Zone/lane partitioning & 3/4 & 4/4 \\
Anti-regression feedback & 3/4 & 2/4 \\
Worked policy examples & 2/4 & 3/4 \\
Cleaning cost economics & 2/4 & 3/4 \\
\textbf{Explicit fairness mechanism} & \textbf{0/4} & \textbf{4/4} \\
\quad \emph{of which:} time-based role rotation & 0/4 & 3/4 \\
\quad \emph{of which:} synchronized clean/collect & 0/4 & 1/4 \\
\bottomrule
\end{tabular}
\end{table}

\begin{figure}[h]
\centering
\includegraphics[width=0.79\columnwidth]{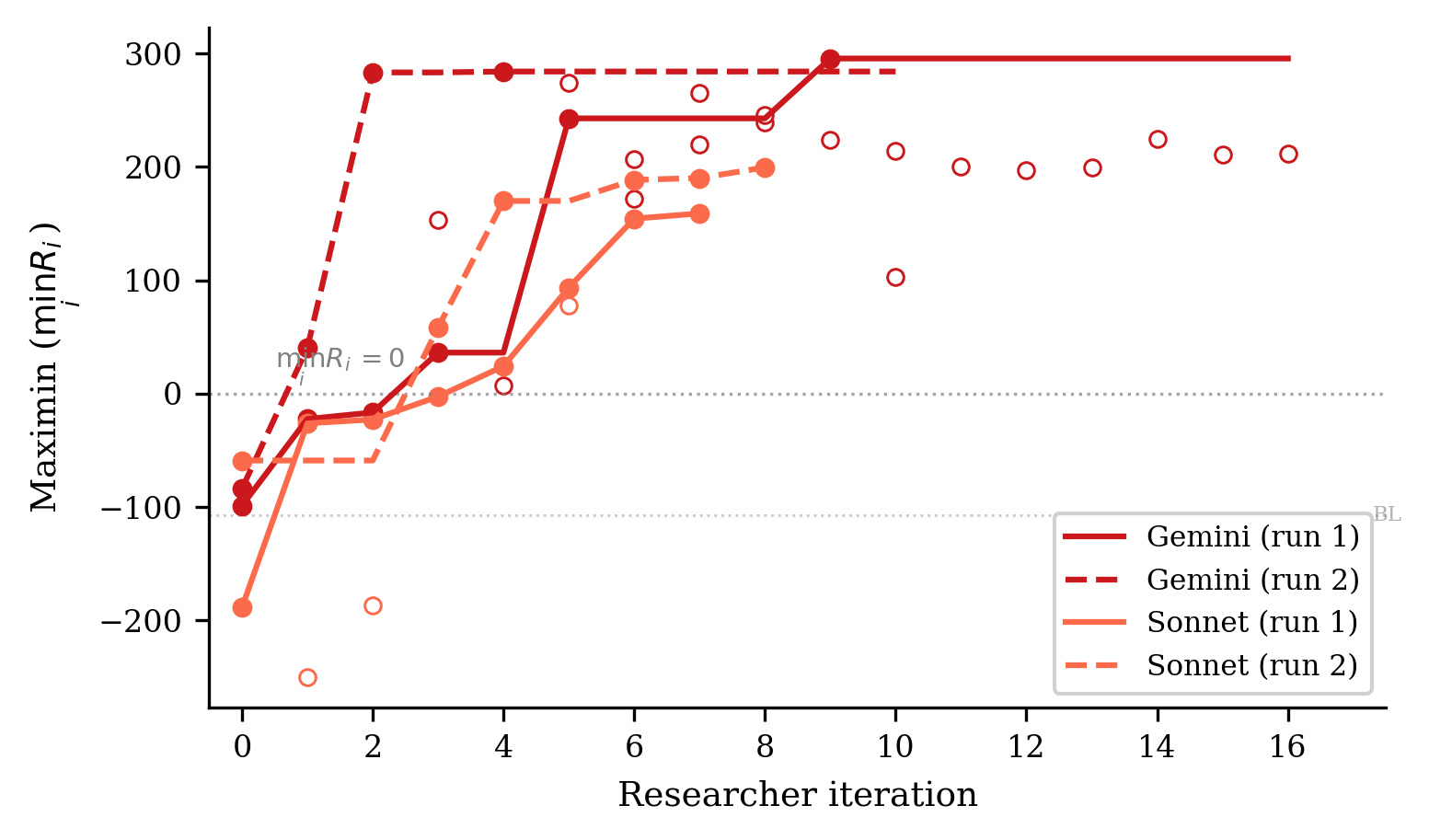}
\caption{Maximin ($\min_i R_i$) across researcher iterations for the 4 Cleanup maximin-optimized runs. All runs transform deeply negative baselines (worst-off agents losing reward) into substantially positive values. Gemini reaches ${\sim}290$ while Sonnet reaches ${\sim}160$--$200$. The dashed line marks $\min_i R_i = 0$ (no agent loses reward overall).}
\label{fig:maximin}
\end{figure}

\begin{figure}[h]
\centering
\includegraphics[width=\columnwidth]{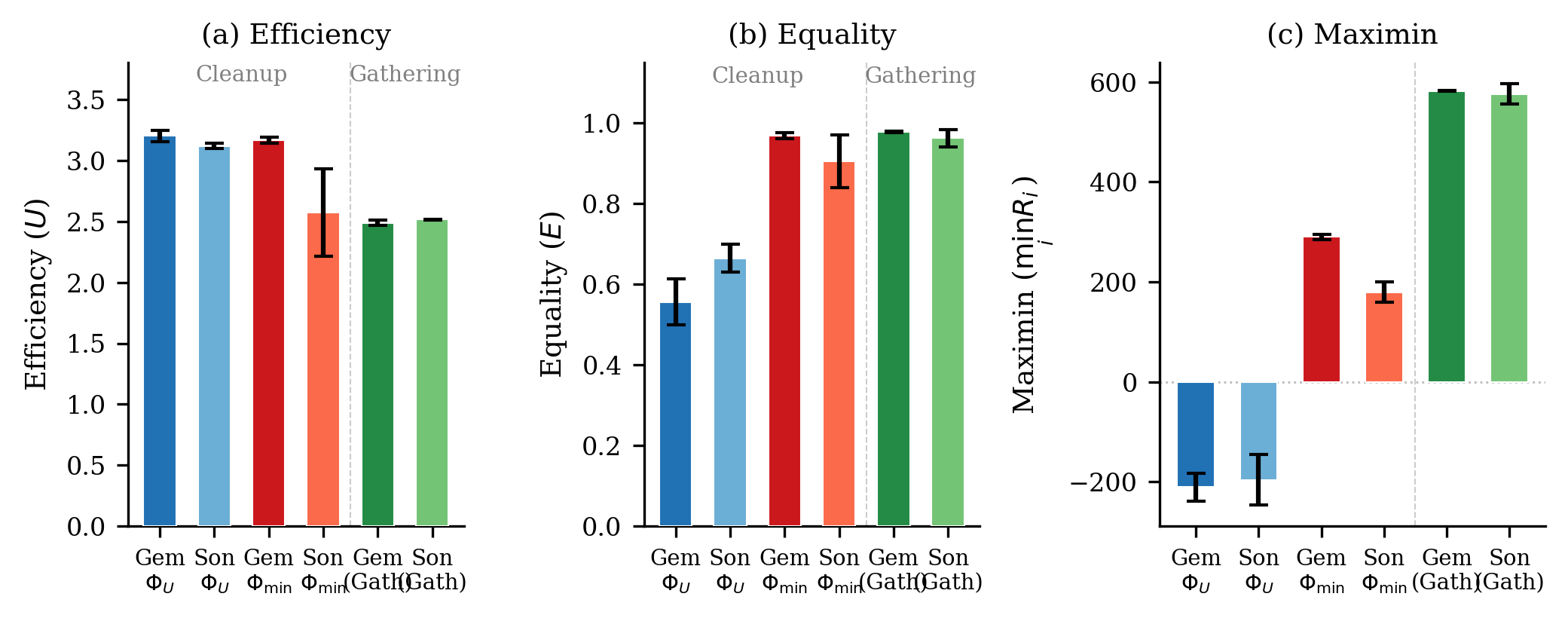}
\caption{Final metrics across all conditions. (a)~Efficiency: all conditions converge to $U \approx 2.5$--$3.2$. (b)~Equality: maximin-optimized Cleanup runs achieve $E \approx 1.0$, while efficiency-optimized runs show $E \approx 0.5$--$0.7$; Gathering achieves high equality regardless. (c)~Maximin: the sharpest contrast---efficiency optimization leaves the worst-off agent deeply negative ($\min_i R_i \approx {-}200$), while maximin optimization transforms it to $+290$ (Gemini) or $+179$ (Sonnet). Gathering achieves high maximin (${\sim}580$) under efficiency optimization alone. Error bars show s.d.\ across runs.}
\label{fig:eff_eq}
\end{figure}

\begin{figure}[h]
\centering
\includegraphics[width=0.79\textwidth]{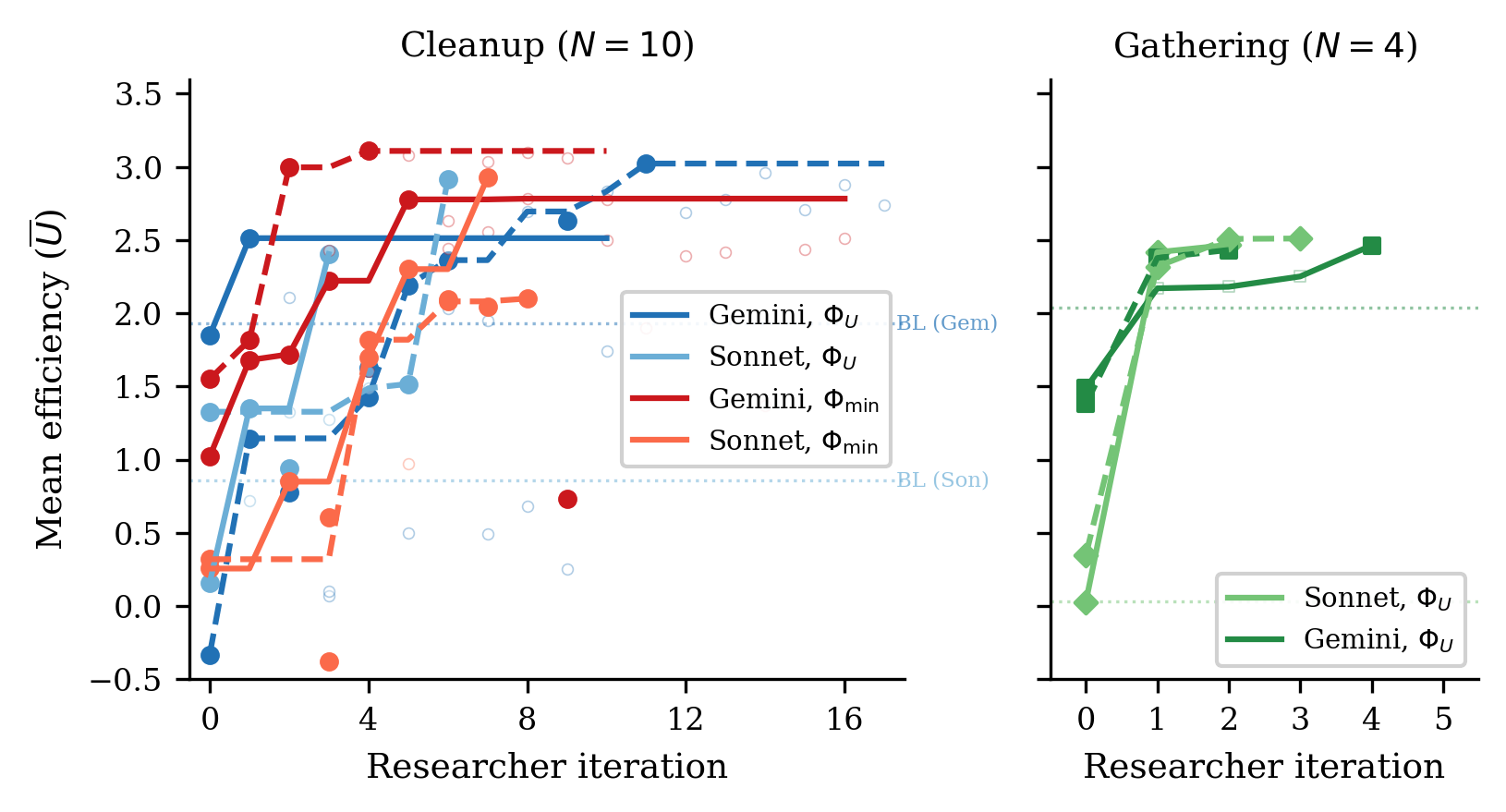}
\caption{Mean efficiency $\bar U$ \emph{averaged across all inner iterations} of each outer step (compare Figure~\ref{fig:efficiency}, which shows the kept policy only). \emph{Left}: Cleanup ($N{=}10$); \emph{right}: Gathering ($N{=}4$). The trajectories track Figure~\ref{fig:efficiency} closely, indicating that the researcher's gains come from improving the entire inner-loop output distribution, not just the best of $K$ samples.}
\label{fig:efficiency_avg}
\end{figure}

\begin{figure}[h]
\centering
\includegraphics[width=0.79\columnwidth]{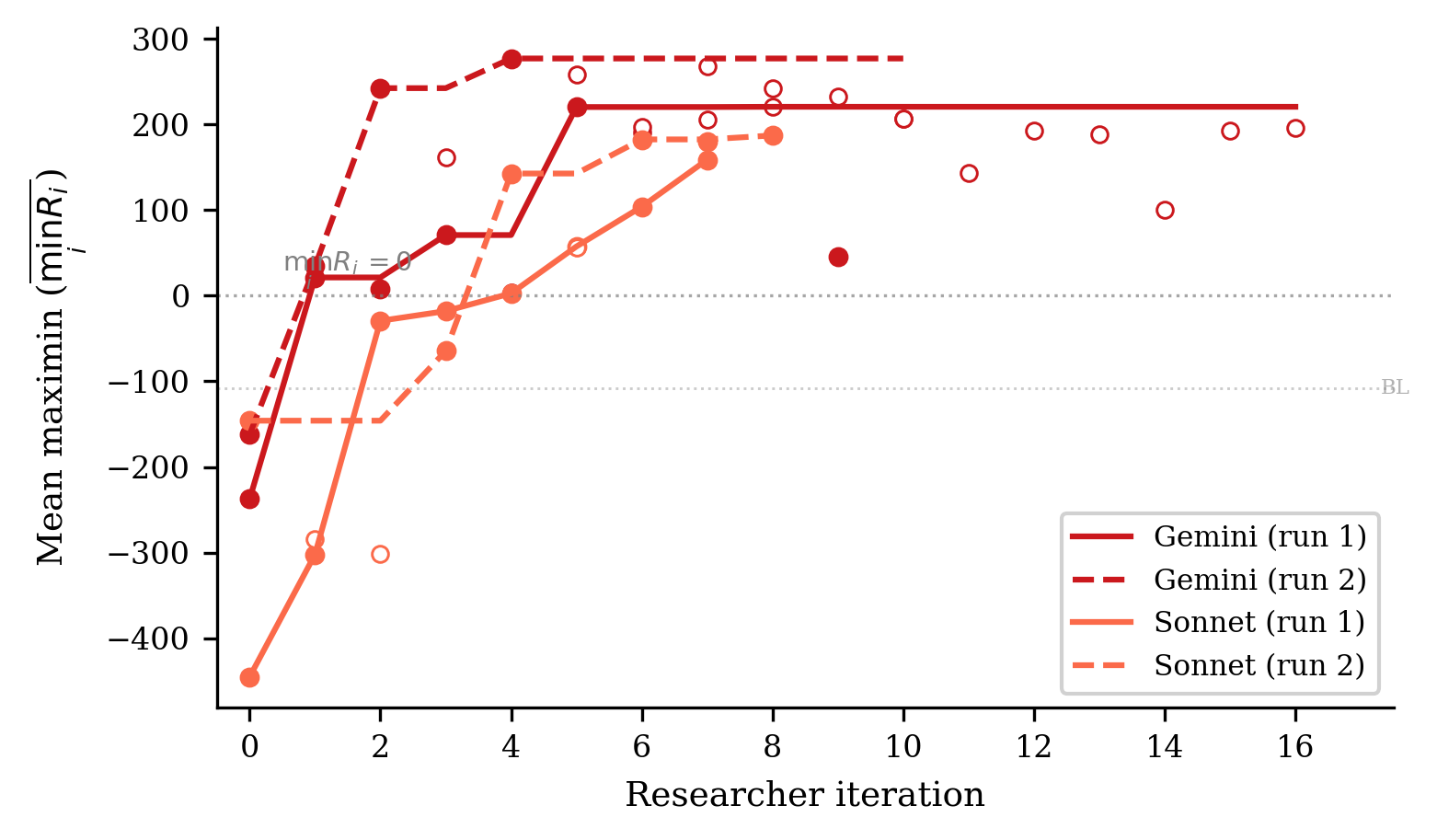}
\caption{Mean maximin $\overline{\min_i R_i}$ \emph{averaged across all inner iterations} of each outer step for the 4 Cleanup $\Phi_{\min}$ runs (compare Figure~\ref{fig:maximin}). The deeply negative baselines lift to $\sim$200--275 mean maximin, only mildly below the kept-policy values of Figure~\ref{fig:maximin}. The horizontal dashed line marks $\overline{\min_i R_i} = 0$.}
\label{fig:maximin_avg}
\end{figure}

\paragraph{Inner-loop averages confirm a broad-based improvement.}
Figures~\ref{fig:efficiency} and~\ref{fig:maximin} report the metric of the \emph{kept} inner-loop output ($\pi^*_c$) at each outer iteration. A natural concern is that this could overstate pipeline quality if the researcher is benefitting from variance: with $K$ inner iterations per outer step, a single lucky generation could carry the curve while the rest of the inner trajectory is noise. Figures~\ref{fig:efficiency_avg} and~\ref{fig:maximin_avg} replot the same runs with the metric averaged across \emph{all} inner iterations of each outer step. The trajectories ramp at essentially the same rate, with only mild attenuation: Cleanup mean efficiency still climbs to $\bar U \approx 2.5$--$3.0$ (vs.\ $3.1$--$3.2$ for the kept policy), Gathering still saturates at $\bar U \approx 2.5$, and Cleanup mean maximin still reaches $\overline{\min_i R_i} \approx 200$--$275$ (vs.\ $179$--$290$). The whole inner-loop output distribution improves, not just its tail, consistent with the interpretation that the researcher is shaping the synthesizer's behavior rather than amplifying lucky draws.

\subsection{Discard Taxonomy}
\label{app:discards}
\begin{table}[!h]
\caption{Discard taxonomy across 12 main-experiment runs (100 outer iterations). The four named-mode categories (iteration regression, over-prescription, feedback overload, variance probe) correspond to the failure patterns described in Section~\ref{sec:experiments}; ``pure $J$-regression'' captures iterations where $\mathcal{R}$'s edit underperformed the running best $J^*$ without an identifiable named cause.}
\label{tab:discards}
\centering
\small
\begin{tabular}{@{}lrr@{}}
\toprule
Category & Count & Share of discards \\
\midrule
Kept ($J_j > J^* + \tau$) & 47 & --- \\
Baselines (unmodified pipeline, $j{=}0$) & 7 & --- \\
\midrule
\textbf{Discarded} ($J_j \le J^*$) & \textbf{46} & \textbf{100\%} \\
\quad pure $J$-regression (no identified failure mode) & 20 & 43\% \\
\quad iteration regression (\textit{e.g.}, $K{\in}\{4,5\}$ over-refinement) & 11 & 24\% \\
\quad over-prescription (verbose hints / restrictive worked examples) & 8 & 17\% \\
\quad feedback overload (verbose diagnostics, anti-regression guards) & 4 & 9\% \\
\quad variance probe (deliberate re-run at fixed $c$ to gauge seed noise) & 3 & 7\% \\
\midrule
Total outer iterations logged & 100 & \\
\bottomrule
\end{tabular}
\end{table}

A natural question raised by Figure~\ref{fig:efficiency} is how much of the open-circle ``discard'' mass is genuine exploration that regresses on $J$ versus iterations that fail for reasons unrelated to the objective. Aggregating across all 12 main-experiment runs (Table~\ref{tab:discards}), $\mathcal{R}$ ran 100 outer iterations: 7 baselines (the unmodified pipeline at $j{=}0$ for each run, plus two re-baselines), 47 kept ($J_j > J^* + \tau$ with $\tau{=}0$), and 46 discarded. No outer iteration was discarded for AST safety or smoke-test failure; those events occur \emph{inside} the inner loop and are absorbed by the regeneration mechanism (up to $R$ retries with the error message appended to the prompt, Section~\ref{sec:framework}). All 46 discards are therefore regressions on $J$; we further classify them by which of the failure modes in Section~\ref{sec:experiments} (``Common failure modes'') we can identify from the researcher's logged description and the per-iteration metric trajectory.

Three observations follow. First, no discards are generation/validation failures at the outer loop, confirming that the AST-and-smoke-test guard plus regeneration-on-error is a non-leaky boundary between inner-loop code-validity issues and outer-loop strategy decisions. Second, the named failure modes from Section~\ref{sec:experiments} account for $23/46$ ($50\%$) of discards; the remaining $20/46$ ($43\%$) are ``pure'' regressions on $J$ where $\mathcal{R}$ tried an intuitively plausible modification (a new helper, a reworded hint, a different $|S|$) that simply underperformed the running best. Third, the researcher occasionally elects to spend an iteration on a deliberate variance probe (re-running the same $c$ on a different seed set) --- these constitute $3/46$ of discards and reflect $\mathcal{R}$ actively reasoning about evaluator noise rather than only chasing $J$. The breakdown by cell is qualitatively consistent: under $\Phi_U$ on Cleanup, iteration regression ($7/22$) and feedback overload ($4/22$) dominate, matching the volatility of high-$U$ Cleanup policies; under $\Phi_{\min}$ on Cleanup, pure $J$-regression dominates ($12/21$), reflecting the broader plateau of competitive maximin configurations $\mathcal{R}$ explores before locking in a rotation template; Gathering is sparser ($3$ discards across $4$ runs), consistent with that game's faster saturation.

\subsection{Spec-gaming inspection}
\label{app:specgaming}

The configuration tuple $c=(p,\phi,\mathcal{H},\iota)$ is unconstrained: $\cR$ is free to edit the helper library $\mathcal{H}$ and the feedback function $\phi$ in ways that could in principle game the held-out evaluation $J$ (e.g., by exposing simulator internals to $\mathcal{M}$, hard-coding actions, or overfitting hyperparameters to the eval seed set). We inspected the final pipelines from all runs for such patterns. We found no helpers that hard-code action sequences, manipulate the environment's RNG, or expose information beyond what is in principle observable from the gridworld state: $\mathcal{H}$'s edits across runs are spatial heuristics (BFS-Voronoi, zone partitioning), state queries (\texttt{waste\_fraction}, \texttt{count\_alive\_apples\_in\_cols}, \texttt{apples\_respawning\_soon}), and action-effect previews (\texttt{best\_clean\_orientation} simulates the beam geometry). $\phi$'s edits are \emph{thresholded interventions on the optimized metric itself} (e.g., a ``\textsc{regression}'' guard when latest $U <$ running-best $U$; a ``\textsc{problem}'' alert when maximin $< 0$): these condition feedback on the primary signal $\cR$ is being scored on, and use thresholds tied to the metric's natural scale rather than to seed-specific values. On the iteration logic side, $\cR$ tends to walk $|S|$ \emph{upward} ($5{\to}8{\to}12$) when chasing maximin, which reduces seed-noise rather than overfitting to a small seed set. The closest behavior to spec-gaming we observed is the over-prescription failure mode of Section~\ref{sec:experiments} (writing such a detailed worked example that $\mathcal{M}$ effectively copies it) which manifests as a \emph{regression} on $J$, not an undeserved gain, because the copied policy is brittle on held-out seeds.

\subsection{Smaller Open-Weight Model: Gemma 4 26B}
\label{app:gemma}

We test the framework with Gemma 4 26B-A4B-IT (Google), a 26B-parameter open-weight model substantially smaller than the frontier models in the main experiments. We run three experiment runs: two on Cleanup optimizing efficiency and maximin respectively, and one on Gathering optimizing efficiency. As with the other synthetizer models, all with Opus 4.6 as the researcher $\cR$.

\paragraph{Setup.} In all three experiments, Gemma starts from complete failure: the baseline pipeline produces $U{=}{-}10.0$ on Cleanup (agents CLEAN-spam without collecting apples) and $U{=}0.0$ on Gathering (broken BFS calls), compared to $U{=}1.93$/$0.86$ (Gemini/Sonnet on Cleanup) and $U{=}2.04$/$0.03$ (Gemini/Sonnet on Gathering). The researcher ran $J_{\max}{=}10$--$18$ outer iterations per condition.

\paragraph{Results.} Table~\ref{tab:gemma} presents the best configurations discovered. Under efficiency optimization, the researcher achieves $U{=}0.87$, a dramatic recovery from the broken baseline but far below frontier models ($U{\approx}3.2$). The researcher compensates for the model's limited capability by reducing inner-loop iterations to $K{=}1$ (avoiding the regression that plagued $K{\geq}2$ runs) and providing highly structured worked examples with explicit cleaning-role assignment.

\begin{table}[h]
\caption{Autoresearch with Gemma 4 26B-A4B-IT. Single run per condition. Baseline: pipeline from~\cite{gallego2026}.}
\label{tab:gemma}
\centering
\small
\begin{tabular}{@{}llcccc@{}}
\toprule
Game & Target & $U$ & $E$ & $\min_i R_i$ & Keep rate \\
\midrule
\multirow{3}{*}{\rotatebox[origin=c]{0}{\scriptsize Cleanup}}
& Baseline & $-$10.0 & 1.00\textsuperscript{$\dagger$} & $-$1000 & --- \\
& $\Phi_U$ & 0.87 & $-$1.11 & $-$371 & 6/19 \\
& $\Phi_{\min}$ & \textbf{1.71} & \textbf{0.94} & \textbf{137} & 9/17 \\
\addlinespace
\multirow{2}{*}{\rotatebox[origin=c]{0}{\scriptsize Gathering}}
& Baseline & 0.0 & 1.00\textsuperscript{$\dagger$} & 0 & --- \\
& $\Phi_U$ & \textbf{2.44} & \textbf{0.98} & \textbf{580} & 4/11 \\
\bottomrule
\multicolumn{6}{@{}l}{\textsuperscript{$\dagger$}\scriptsize Equality is trivially 1.0 because all agents receive near-zero reward.}
\end{tabular}
\end{table}

\paragraph{Maximin optimization rescues efficiency.}
Strikingly, under maximin optimization the researcher achieves \emph{higher} efficiency ($U{=}1.71$) than under direct efficiency optimization ($U{=}0.87$), alongside strong equality ($E{=}0.94$) and positive worst-off welfare ($\min_i R_i{=}137$). This reversal, absent in frontier models where both objectives yield $U{\approx}3.1$--$3.2$, occurs because the maximin objective forces the researcher toward coordination mechanisms (rotating cleaning duties, index-based apple assignment) that simultaneously improve collective output. Under efficiency-only optimization, the researcher converges on a local optimum that a 26B model can implement but that caps performance well below the frontier.

This suggests that \emph{for models below a capability threshold, fairness objectives may serve as better optimization targets for overall social welfare than direct efficiency maximization}. The structured coordination enforced by the maximin objective provides a scaffold that compensates for the weaker model's difficulty implementing complex strategies from strategic hints alone.

\paragraph{Gathering: full recovery on a simpler game.}
On Gathering ($N{=}4$), the researcher brings Gemma from $U{=}0.0$ to $U{=}2.44$ with $E{=}0.98$ and $\min_i R_i{=}580$, after 10 outer iterations (4 kept). These values nearly match frontier models (Gemini: $U{=}2.49$, Sonnet: $U{=}2.52$; Table~\ref{tab:results}). The researcher discovers the same Voronoi partitioning and respawn-aware camping strategies found in the main experiments, and increases inner-loop iterations to $K{=}5$ to allow sufficient refinement. The contrast with Cleanup suggests that the researcher can fully compensate for model weakness on simpler coordination tasks (4 agents, symmetric costs) but not on harder provision dilemmas (10 agents, asymmetric costs).

\subsection{Compute Requirements}
\label{app:compute}

Table~\ref{tab:compute} reports wall-clock time for all 12 autoresearch runs. \emph{Inner loop} measures total time spent on policy LLM generation and environment simulation across all evaluations; \emph{total} (estimated from run-directory timestamps) additionally includes the researcher agent's analysis, code editing, and decision-making between evaluations.

\begin{table}[h]
\caption{Wall-clock time per autonomous run. Evals: total inner loop evaluations including initial baseline. Per eval: mean wall-clock per single evaluation. Total: end-to-end including researcher overhead.}
\label{tab:compute}
\centering
\footnotesize
\begin{tabular}{@{}llcrrr@{}}
\toprule
Policy LLM & $\Phi$ & Evals & \shortstack{Inner\\loop (h)} & \shortstack{Per\\eval (min)} & \shortstack{Total\\(h)} \\
\midrule
\multicolumn{6}{@{}l}{\emph{Cleanup ($N{=}10$)}} \\
Gemini & $\Phi_U$ & 11 & 2.6 & 14 & 3.7 \\
Gemini & $\Phi_U$ & 18 & 4.1 & 14 & 6.4 \\
Sonnet & $\Phi_U$ & 4 & 2.1 & 32 & 6.1 \\
Sonnet & $\Phi_U$ & 7 & 5.0 & 43 & 6.1 \\
Gemini & $\Phi_{\min}$ & 17 & 3.0 & 11 & 4.3 \\
Gemini & $\Phi_{\min}$ & 11 & 3.6 & 20 & 5.0 \\
Sonnet & $\Phi_{\min}$ & 8 & 4.5 & 33 & 8.8 \\
Sonnet & $\Phi_{\min}$ & 9 & 5.3 & 36 & 13.2 \\
\addlinespace
\multicolumn{6}{@{}l}{\emph{Gathering ($N{=}4$)}} \\
Sonnet & $\Phi_U$ & 3 & 1.7 & 33 & 2.2 \\
Gemini & $\Phi_U$ & 5 & 1.4 & 17 & 1.9 \\
Gemini & $\Phi_U$ & 3 & 0.9 & 19 & 1.1 \\
Sonnet & $\Phi_U$ & 4 & 2.3 & 35 & 3.4 \\
\midrule
\multicolumn{2}{@{}l}{All 12 runs} & 100 & 36.6 & 22 & 62.2 \\
\bottomrule
\end{tabular}
\end{table}

\paragraph{Cost breakdown.}
Each inner loop evaluation involves $K{=}2$--$3$ policy LLM generation calls (each ${\sim}$2k input tokens, ${\sim}$1.5k output tokens) followed by multi-seed simulation (5--12 seeds, ${\sim}$15--30s total). Policy LLM generation dominates inner loop time (86--97\%), with Sonnet evaluations taking ${\sim}2{\times}$ longer than Gemini due to extended thinking. The researcher agent (Claude Opus 4.6, running via Claude Code CLI) accounts for 41\% of total wall-clock time (25.6h of the 62.2h total across all 12 runs), spent reading results, editing pipeline source files, and planning modifications. In monetary terms, the researcher agent is the dominant cost: each outer iteration consumes ${\sim}$50--100k context tokens in the Opus session, whereas each inner loop evaluation uses only ${\sim}$5--10k policy LLM tokens.

\section{Researcher-Authored Pipeline Artifacts}
\label{app:artifacts}

The previous appendix shows the policies $\pi^*_c$ that the inner loop \emph{outputs}; this one shows the configuration $c = (p, \phi, \mathcal{H}, \iota)$ (Section~\ref{sec:experiments}, Table~\ref{tab:config}) that the researcher $\cR$ \emph{authored} to produce them. Excerpts are taken from the final commit on the dedicated git branch of the corresponding run; we reproduce the prose verbatim, with author commentary in \texttt{[brackets]} and elisions marked \texttt{...}. We organize by artifact type so each subsection makes a within-type contrast (e.g., $\Phi_U$ vs.\ $\Phi_{\min}$, or Cleanup vs.\ Gathering). Subsection~\ref{app:art-r-prompt} comes first because it shows the upstream input that the rest of the appendix is derived from: the prompt $p_\cR$ given to the researcher itself.

\subsection{Researcher system prompt $p_\cR$}
\label{app:art-r-prompt}

The researcher $\cR$ is a coding agent (Claude Opus 4.6 via the Claude Code CLI) instantiated with a single fixed system prompt that is identical across all experiment runs (it differs only in the game name and the primary-metric flag, which is one of \texttt{efficiency} or \texttt{maximin}). This is the input from which every artifact in the rest of this appendix is derived. Because much of the paper's claim hinges on the asymmetry between \emph{what $p_\cR$ tells $\cR$} and \emph{what $\cR$ in turn writes for $\cM$}, we reproduce the strategically relevant portions of $p_\cR$ verbatim below; the full file is in the released repository under \texttt{autoresearch/program.md}.

\begin{lstlisting}[style=prompttext, caption={Excerpt from $p_\cR$. The remaining parts of $p_\cR$ describe the file layout of the inner-loop pipeline, the evaluation script invocation, the keep/discard rule, and the logging format; none of those add task-specific strategic guidance.}, label={lst:r-prompt}]
## The metrics

The primary metric is specified at launch via `--metric` (default: `efficiency`).
The two options are:

**Efficiency** (U): collective apple collection rate across all agents per
timestep. Higher is better. ...

**Maximin** (Rawlsian welfare): minimum total per-agent return across all
agents. Higher is better. Inspired by Rawls' difference principle -- a just
policy maximizes the welfare of the worst-off agent. ...

## Strategy space

Here are categories of modifications to explore:

### Prompt engineering (p)
- Add strategic hints about the Cleanup dilemma (e.g., "cleaning is a public
  good -- someone must do it")
- Add worked examples of sophisticated policies
- Restructure the API documentation for clarity
- Add game-theoretic reasoning frameworks
- Mention optimal strategies from the literature (Voronoi partitioning,
  role assignment)

### Feedback engineering (l, phi)
- Show per-agent reward breakdown (not just average)
- Add derived metrics (e.g., cleaning rate, waste level trends, apple growth)
- Frame feedback to emphasize cooperation
- Add temporal analysis ...
- Show metric trends across iterations ...
- Provide diagnostic hints based on metrics ...

### Helper functions (H)
- count_waste(env), waste_fraction(env), bfs_to_waste(env, agent_id),
  should_clean(env), assign_role(env, agent_id),
  find_cleaning_position(env, agent_id)

### Iteration logic (iota)
- Change K (more iterations = more refinement but more cost)
- Change eval seeds, retry budget, thinking budget
\end{lstlisting}

\paragraph{What $p_\cR$ does \emph{not} say.}
For the convergent-discovery claim of Finding~4 to be informative, $p_\cR$ must not itself encode the specific mechanisms that $\cR$ later writes into $p$, $\phi$, $\mathcal{H}$ under $\Phi_{\min}$. Inspecting Listing~\ref{lst:r-prompt}, three absences matter:
\begin{itemize}
  \item \emph{No rotation formula.} $p_\cR$ never mentions \texttt{(agent\_id + env.\_step\_count // T) \% n}, \texttt{env.\_step\_count}, or any other time-based cycling pattern. The closest item -- ``Mention optimal strategies from the literature (Voronoi partitioning, role assignment)'' -- names \emph{static} role assignment, not time-rotated duty.
  \item \emph{No objective-conditional guidance.} The strategy-space listing is identical regardless of whether $\cR$ is launched with \texttt{--metric efficiency} or \texttt{--metric maximin}. $p_\cR$ does not tell $\cR$ to behave differently under the two objectives; the only objective-sensitive input is the (scalar) score $J_j$ returned after each inner-loop run.
  \item \emph{No ``rotation is for fairness.''} $p_\cR$ does not link any strategy to either welfare criterion. The connection ``rotation = fairness = high maximin'' is constructed by $\cR$ from $J_j$ observations across iterations on the branch.
\end{itemize}
The researcher-authored prompt $p$ shown in Listing~\ref{lst:prompt-min} and its Sonnet counterpart, both written under $\Phi_{\min}$, contain all three of these elements; the corresponding $\Phi_U$ prompts do not. The role of $p_\cR$ in the experiment is therefore to define the configuration space and the evaluation harness, not to script the discoveries themselves.

\subsection{Helper library $\mathcal{H}$: coordination primitives}
\label{app:art-helpers}

The researcher adds primitives that the policy LLM can call as black boxes, sparing it from re-implementing tricky logic on every iteration. Two patterns dominate: (i)~\emph{state inspection} (waste counts, fractions, beam-yield scoring) and (ii)~\emph{coordination primitives} (zone assignment, role rotation). We show one of each.

\begin{lstlisting}[style=pythonpolicy, caption={Cleanup, $\Phi_{\min}$ -- band-based apple zoning helper added by the researcher in exp5. This is the primitive that lets the policy in Listing~\ref{lst:min-gemini} keep collectors within their own row band, preventing all 7 gatherers from racing to the same apple.}, label={lst:helper-zone}]
def get_my_apples(env, agent_id):
    """Alive apples in this agent's horizontal row band.
    Divides the orchard into n_agents bands; falls back to all apples if empty."""
    aid    = int(agent_id)
    n      = int(env.n_agents)
    band_h = env.height / n
    band_lo, band_hi = aid * band_h, (aid + 1) * band_h

    my_apples, all_apples = set(), set()
    for i in range(env.n_apples):
        if env.apple_alive[i]:
            ar = int(env._apple_pos[i, 0]); ac = int(env._apple_pos[i, 1])
            all_apples.add((ar, ac))
            if band_lo <= ar < band_hi:
                my_apples.add((ar, ac))
    return my_apples if my_apples else all_apples
\end{lstlisting}

\begin{lstlisting}[style=pythonpolicy, caption={Gathering -- BFS-Voronoi territory + respawn-aware waiting helpers added by the researcher in gather-exp3. These are the primitives that Listing~\ref{lst:gather-gemini} composes into a one-line policy: ``go to my\_zone\_apples, otherwise nearest\_respawning\_apple.''}, label={lst:helper-vor}]
def voronoi_zones(env):
    """Multi-source BFS Voronoi over walkable cells.
    Returns (row, col) -> agent_id; ties broken by lower agent_id."""
    queue, visited = deque(), {}
    for a_id in range(env.n_agents):
        if int(env.agent_timeout[a_id]) == 0:
            ar = int(env.agent_pos[a_id][0]); ac = int(env.agent_pos[a_id][1])
            queue.append((ar, ac, a_id))
            visited[(ar, ac)] = a_id
    while queue:
        r, c, a_id = queue.popleft()
        for dr, dc in [(-1,0),(1,0),(0,-1),(0,1)]:
            nr, nc = r + dr, c + dc
            if 0 <= nr < env.height and 0 <= nc < env.width \
                    and not env.walls[nr, nc] and (nr, nc) not in visited:
                visited[(nr, nc)] = a_id
                queue.append((nr, nc, a_id))
    return visited

def nearest_respawning_apple(env, agent_id, zones=None, max_timer=10):
    """Nearest dead apple in MY zone respawning within max_timer steps.
    Returns (row, col) of best wait spot, or None."""
    if zones is None: zones = voronoi_zones(env)
    best_pos, best_t, best_d = None, max_timer + 1, float('inf')
    ar = int(env.agent_pos[agent_id][0]); ac = int(env.agent_pos[agent_id][1])
    for i in range(env.n_apples):
        if not env.apple_alive[i]:
            pos = (int(env._apple_pos[i][0]), int(env._apple_pos[i][1]))
            if zones.get(pos) == agent_id:
                t = int(env.apple_timer[i])
                if t <= max_timer:
                    d = abs(pos[0] - ar) + abs(pos[1] - ac)
                    if t < best_t or (t == best_t and d < best_d):
                        best_pos, best_t, best_d = pos, t, d
    return best_pos
\end{lstlisting}

The choice of helper depends on the dilemma's geometry: row-bands suffice when fairness is the main concern (Cleanup maximin) but full BFS-Voronoi is needed when wall-aware territory ownership matters (Gathering). The researcher does not fix this in advance, it picks the simpler primitive whenever it works.

\subsection{System prompt $p$: from neutral framing to strategic briefing}
\label{app:art-prompts}

The unmodified system prompt is a neutral API description: \emph{``Write a policy that maximizes per-agent reward.''} Under maximin optimization, the researcher rewrites the prompt into a multi-section strategic briefing. The excerpt below shows the pieces it added to the Cleanup prompt in exp5; the full prompt grew from 165 to 325 lines.

\begin{lstlisting}[style=prompttext, caption={Cleanup, $\Phi_{\min}$ -- excerpts from the researcher-rewritten system prompt (exp5). The unmodified baseline contained only API documentation; everything below was added by $\cR$.}, label={lst:prompt-min}]
## CRITICAL OBJECTIVE: Rawlsian Fairness (Maximin)

Your goal is to **maximize the minimum per-agent total return** across all
agents. This is the "maximin" or Rawlsian welfare criterion.

This means: it is NOT enough for the *average* agent to do well. The
**worst-off agent** must do as well as possible. A policy where 8 agents
each earn +200 but 2 agents each earn -100 is TERRIBLE (maximin = -100).

**Key implication**: cleaning costs (-1 per CLEAN action) must be shared
equitably among ALL agents. If you assign fixed "cleaner" roles, those
agents will accumulate large negative rewards and destroy your maximin score.

## Strategy for Maximin: ROLE ROTATION + APPLE ZONING

The optimal strategy for maximin involves:
1. **Shared cleaning duty**: ALL agents take turns cleaning using a rotation
   schedule based on `agent_id` and `env._step_count`. For example:
   `is_my_cleaning_turn = (agent_id + env._step_count // SHIFT) % env.n_agents < NUM_CLEANERS`
   where SHIFT is ~50 steps and NUM_CLEANERS is 2-3.
2. **When NOT your turn**: collect apples from YOUR ZONE using
   `get_my_apples(env, agent_id)` -- prevents all gatherers competing
   for the same nearest apple.
3. **NEVER use BEAM (action 6)**: it costs -1 and causes -50 to the target.
4. **Keep waste density low**: aim for ~2-3 active cleaners at any time.

[... API documentation, helper documentation, working example ...]

IMPORTANT:
- NEVER assign permanent cleaning roles to specific agents -- this kills maximin.
- Use env._step_count for time-based role rotation so all agents share cleaning duty.
- NEVER use BEAM (action 6) -- it destroys both agents' rewards.
\end{lstlisting}

Two things stand out. First, the researcher writes the formula it expects $\cM$ to use almost verbatim ( \texttt{(agent\_id + env.\_step\_count // SHIFT) \% env.n\_agents < NUM\_CLEANERS} ) and the policy in Listing~\ref{lst:min-gemini} uses essentially this template. Second, the researcher \emph{teaches by counter-example}: the explicit ``8 agents earn +200, 2 earn -100, maximin = -100'' makes the failure mode of $\Phi_U$-style static roles (Listing~\ref{lst:eff-gemini}) concretely visible to $\cM$. This is the information-design move predicted by the mechanism-design framing in Section~\ref{sec:mechdesign}.

The Gathering prompt evolves along a different axis (no maximin, no rotation). Its researcher-added ``Key Strategic Insights'' section instead emphasizes \emph{(i)~self-play implies never beam} and \emph{(ii)~Voronoi partition each step}, exactly matching what the policy in Listing~\ref{lst:gather-gemini} implements.

\subsection{Feedback $\phi$: adaptive diagnostics}
\label{app:art-feedback}

The baseline feedback function from~\cite{gallego2026} shows reward + four social metrics with definitions and stops there. The researcher turns it into a state machine: it inspects the latest metrics and conditionally injects different hints. Two different runs produced two qualitatively different diagnostics: the same metric ($\bar r_k$ or $\min_i R_i$) is repurposed to either \emph{stabilize} a working policy or \emph{redirect} a failing one.

\begin{lstlisting}[style=pythonpolicy, caption={Adaptive feedback diagnostics. \emph{Top}: efficiency-optimized run (exp1) -- a stability guard prevents iter-3 regression once $U$ is high. \emph{Bottom}: maximin-optimized run (exp5) -- two fairness diagnostics that fire when the worst-off agent loses reward.}, label={lst:feedback-adapt}]
# --- Cleanup, Phi_U feedback (exp1, pipeline/feedback.py final) ---
last_eff = history[-1]["metrics"]["efficiency"]
if last_eff >= 2.5:
    parts.append(
        "**CRITICAL -- DO NOT REGRESS**: The current policy achieves high "
        "efficiency (>=2.5). You MUST output a policy that is nearly identical "
        "to the current one. Copy the current policy code and make AT MOST "
        "one small targeted improvement (e.g., adjust a single numeric "
        "threshold). If you are not confident a change will help, output the "
        "current policy UNCHANGED. A regression here means the run fails.")

# --- Cleanup, Phi_min feedback (exp5, pipeline/feedback.py final) ---
last_maximin = history[-1]["metrics"].get("maximin", 0)
last_avg     = history[-1]["reward_avg"]
if last_maximin < 0:
    parts.append(
        f"**FAIRNESS ALERT**: maximin={last_maximin:.1f} is NEGATIVE. "
        f"The worst-off agent lost reward overall while average was {last_avg:.1f}. "
        "This means cleaning duties are NOT shared equitably. "
        "Use time-based role rotation (env._step_count) so ALL agents share "
        "cleaning costs equally. NEVER assign permanent cleaner roles.")
elif last_maximin < last_avg * 0.5:
    parts.append(
        f"**FAIRNESS WARNING**: maximin={last_maximin:.1f} is much lower than "
        f"average={last_avg:.1f}. The gap suggests unequal cleaning burden. "
        "Ensure ALL agents rotate through cleaning duty.")
\end{lstlisting}

The two diagnostics are written by independent researcher runs but converge on a common pattern: \emph{thresholded interventions on a primary metric}. They are also a common cause of the low run-to-run variance reported in Finding~1: when a working policy lands in the high-$U$ basin, the stability guard prevents the LLM from over-refining and tipping out of it (the $K{=}3$ regressions visible in Section~\ref{sec:experiments}'s ``Common failure modes'' (4)), and when an unfair policy lands in the negative-maximin region, the alert injects exactly the rotation hint that recovers it.

\subsection{Iteration logic $\iota$: per-condition hyperparameters}
\label{app:art-config}

Although $\iota$ has the smallest source surface, the researcher's edits to it explain a meaningful fraction of the variance reduction noted in Finding~1. Table~\ref{tab:iota} summarizes the values shipped in the final commit of each best-performing run.

\begin{table}[h]
\caption{Iteration-logic ($\iota$) values in the final commit of selected runs. $K$ = inner-loop iterations, $|S|$ = evaluation seeds, ``thinking'' = extended-thinking token budget passed to $\cM$. Baseline (\cite{gallego2026}) is $K{=}3$, $|S|{=}5$, thinking $16$k.}
\label{tab:iota}
\centering
\footnotesize
\begin{tabular}{@{}llccc l@{}}
\toprule
Run & Game / objective & $K$ & $|S|$ & thinking & Researcher's stated rationale \\
\midrule
exp1 (Gemini)  & Cleanup, $\Phi_U$    & 3 & 5  & 16k & default; stability comes from feedback hint \\
exp4 (Sonnet)  & Cleanup, $\Phi_U$    & 3 & 5  & 32k & ``Sonnet was over-refining at 16k thinking'' \\
exp5 (Gemini)  & Cleanup, $\Phi_{\min}$ & \textbf{2} & \textbf{12} & 16k & ``$K{=}2$ avoids iter-3 regression; 12 seeds for variance control'' \\
exp7 (Sonnet)  & Cleanup, $\Phi_{\min}$ & 3 & 5  & \textbf{10k} & ``reduced thinking from 16k -- Sonnet was generating runaway code at 16k'' \\
gather-exp3 (Gemini) & Gathering, $\Phi_U$ & 3 & 5  & 16k & default; Gathering converges in $K{\le}3$ \\
\bottomrule
\end{tabular}
\end{table}

The maximin Gemini run is the most informative case: the researcher discovered that with $K{=}3$ and $|S|{=}5$, identical configurations could yield $\min_i R_i {=} 295$ on one set of seeds and $\min_i R_i {=} 100$ on another (this is the ``variance exposure'' failure mode of Section~\ref{sec:experiments}'s common failures). It then walked $|S|$ from $5 \to 8 \to 12$ over three outer iterations until the seed-averaged signal was stable enough that the policy LLM stopped chasing noise. The policy in Listing~\ref{lst:min-gemini} is sampled at $|S|{=}12$.

\subsection{Cross-artifact observations}
\label{app:art-cross}

\begin{itemize}
  \item \emph{Helpers do the algebra; prompts pick the strategy class.} The researcher uses helpers (\ref{lst:helper-zone},~\ref{lst:helper-vor}) to put non-trivial primitives at $\cM$'s fingertips, and the prompt (\ref{lst:prompt-min}) to tell $\cM$ \emph{which} primitive to compose. Prompts without supporting helpers produced policies that ``mention rotation'' but failed to implement it; helpers without prompt updates often went unused.
  \item \emph{Feedback is the only adaptive component.} $p$, $\mathcal{H}$, and $\iota$ are static within a run; $\phi$ is the only place where the researcher gets to change what $\cM$ sees \emph{during} the inner loop, and it uses thresholded diagnostics (Listing~\ref{lst:feedback-adapt}) to do so. This is what mostly drives the run-to-run variance reduction.
  \item \emph{The same artifact has different content under each objective.} The Cleanup prompt under $\Phi_U$ omits ``rotation'' and ``maximin'' entirely; the same prompt under $\Phi_{\min}$ devotes its entire opening section to it. The configuration $c$ is genuinely a function of the welfare objective $\Phi$, not a fixed pipeline parameterized by it. This is consistent with the information-design framing of Section~\ref{sec:mechdesign}.
  \item \emph{The researcher's edits are short and targeted.} The full diff between the baseline pipeline and the best maximin Gemini configuration totals ${\sim}300$ added lines spread over four files; ${\sim}80$ of them appear in the listings above. The remaining content is API documentation and worked examples, included for completeness but not strategically novel.
\end{itemize}

\section{Selected Generated Policies}
\label{app:policies}

The code excerpts below are verbatim outputs of the policy synthesizer $\cM$, taken from the final inner-loop iteration of the best-performing run in each condition. To fit the page, we elide repeated boilerplate (BFS scaffolding, fallback branches) with \texttt{...} and add commentary in \texttt{[brackets]}; nothing else has been edited. The listings illustrate the qualitative differences highlighted in Findings 2--4 of Section~\ref{sec:experiments}: \emph{static} vs.\ \emph{rotating} cleaning roles in Cleanup, and the \emph{Voronoi + respawn-timer} structure in Gathering. All four policies were independently generated by the LLM under different seeds, models, and outer-loop runs.

\subsection{Cleanup, $\Phi_U$ (Gemini): static interleaved roles + dynamic threshold}
\label{app:policy-eff-gemini}

Best run: $U{=}3.25$, $E{=}0.61$, $\min_i R_i{=}-182$. The synthesizer assigns \emph{permanent} cleaning roles via $\texttt{agent\_id} \bmod n {<} \tau$, where the threshold $\tau$ depends smoothly on the current waste fraction. Higher pollution recruits more cleaners, but the \emph{identity} of each cleaner is fixed: agents 0--$\tau{-}1$ always pay the cost. This drives $U$ to its ceiling but leaves equality at $0.61$ and pushes the worst-off agent to $\min_i R_i{=}{-}182$.

\begin{lstlisting}[style=pythonpolicy, caption={Cleanup, $\Phi_U$ -- Gemini 3.1 Pro. Static role assignment with waste-tiered threshold (verbatim, abridged).}, label={lst:eff-gemini}]
def policy(env, agent_id) -> int:
    if int(env.agent_timeout[agent_id]) > 0:
        return 7  # STAND while removed

    waste_frac = waste_fraction(env)
    n_agents = env.n_agents

    # [Smooth dynamic threshold: more pollution -> more cleaners]
    if   waste_frac > 0.35: threshold_cleaners = n_agents
    elif waste_frac > 0.25: threshold_cleaners = int(n_agents * 0.7)
    elif waste_frac > 0.15: threshold_cleaners = int(n_agents * 0.5)
    elif waste_frac > 0.05: threshold_cleaners = int(n_agents * 0.3)
    else:                   threshold_cleaners = max(1, int(n_agents * 0.2))

    # [STATIC role: cleaners are ALWAYS the lowest-id agents -- no rotation]
    is_cleaner = (agent_id % n_agents) < threshold_cleaners

    if is_cleaner:
        best_orient, n_waste = find_best_clean_orientation(env, agent_id)
        if n_waste >= 1:
            cur_orient = int(env.agent_orient[agent_id])
            if best_orient == cur_orient: return 8  # CLEAN
            if (cur_orient + 1) % 4 == best_orient: return 5  # ROTATE_RIGHT
            elif (cur_orient - 1) % 4 == best_orient: return 4  # ROTATE_LEFT
            else: return 4
        # [Competitor-aware BFS toward waste in own row band]
        c_idx = agent_id % n_agents
        target_r = int(((c_idx + 0.5) / threshold_cleaners) * env.height)
        # ... pick waste cell minimizing dist + |row - target_r| + 50 * #closer_cleaners
        # ... bfs_toward and direction_to_action
    else:
        # GATHERER: row-banded sweep, competitor-aware
        # assigned_row = ((g_idx + 0.5) / n_gatherers) * env.height
        # ... pick alive apple minimizing dist + |row - assigned_row| + 50 * #closer_gatherers
        ...
\end{lstlisting}

\subsection{Cleanup, $\Phi_{\min}$ (Gemini): time-rotated, geographically interleaved roles}
\label{app:policy-min-gemini}

Best run: $U{=}3.19$, $E{=}0.98$, $\min_i R_i{=}296$. Switching the outer objective from $U$ to $\min_i R_i$ flips the role-assignment scheme entirely: \texttt{role\_idx} now depends on $\texttt{agent\_id} + \texttt{step}/{50}$, so every agent rotates through cleaner and gatherer duty in a 50-step cycle. The cleaner indices ${\{1,4,8\}}$ are spatially \emph{interleaved} along the river (rather than contiguous) so that whoever is on duty has a short walk to their assigned river slice. This single change accounts for the $E{=}0.55 \to 0.98$ jump and the $\min_i R_i {=} {-}182 \to 296$ swing reported in Table~\ref{tab:results}, with negligible efficiency loss.

\begin{lstlisting}[style=pythonpolicy, caption={Cleanup, $\Phi_{\min}$ -- Gemini 3.1 Pro. Time-rotated roles with interleaved geography (verbatim, abridged).}, label={lst:min-gemini}]
def policy(env, agent_id) -> int:
    if int(env.agent_timeout[agent_id]) > 0:
        return 7  # STAND while removed

    aid  = int(agent_id)
    step = int(env._step_count)
    n    = int(env.n_agents)

    # [TIME ROTATION: every 50 steps each agent's role index advances]
    if n == 10:
        role_idx = (aid + step // 50) % 10
        # [Cleaner indices {1,4,8} are SPATIALLY INTERLEAVED, not contiguous,
        #  so each on-duty cleaner has a short walk to its river slice]
        if   role_idx == 1: role_type, zone_idx = 'C', 0
        elif role_idx == 4: role_type, zone_idx = 'C', 1
        elif role_idx == 8: role_type, zone_idx = 'C', 2
        else:
            role_type = 'G'
            g_map = {0:0, 2:1, 3:2, 5:3, 6:4, 7:5, 9:6}
            zone_idx = g_map[role_idx]
        num_cleaners, num_gatherers = 3, 7

    if role_type == 'C':
        # [Each on-duty cleaner owns one horizontal slice of the river]
        row_min = int( zone_idx      * env.height / num_cleaners)
        row_max = int((zone_idx + 1) * env.height / num_cleaners) - 1

        # [Score every (row, col, orientation) in the slice by beam yield;
        #  fire if facing >=2 waste, else move to a >=2-yield position]
        best_positions, good_positions = set(), set()
        for r in range(row_min, row_max + 1):
            for c in range(10):
                if env.walls[r, c] or env.waste[r, c]: continue
                y_max = max(get_clean_yield(env, r, c, o) for o in range(4))
                if y_max >= 2: best_positions.add((r, c))
                if y_max >= 1: good_positions.add((r, c))
        # ... rotate / CLEAN / bfs_to_target_set as appropriate
    else:
        # GATHERER: collect apples ONLY inside the rotating row band
        # row_min/row_max for num_gatherers slices ...
        ...
\end{lstlisting}

\subsection{Cleanup, $\Phi_{\min}$ (Sonnet): independent rediscovery of duty rotation}
\label{app:policy-min-sonnet}

This run was launched on a separate dedicated git branch from the Gemini maximin runs of Listing~\ref{lst:min-gemini} and from $p_\cR$ identical to the one shown in Appendix~\ref{app:art-r-prompt}, with no shared state. The researcher-authored synthesizer prompt $p$ on this branch ended up containing a rotation hint structurally similar to Listing~\ref{lst:prompt-min} but with different phrasing and a different recommended period (\texttt{phase\_length}${\,=\,}100$ in a worked example, vs.\ \texttt{SHIFT}${\,\approx\,}50$ in Listing~\ref{lst:prompt-min}); $\cM$ then chose its own period of $50$. The convergence is therefore at the level of which artifacts $\cR$ injects into $p$, not at the level of $\cM$ inventing rotation from a neutral prompt.

Best run: $U{=}2.93$, $E{=}0.83$, $\min_i R_i{=}154$. The Sonnet synthesizer arrives at the \emph{same} structural insight as the Gemini maximin run: a phase counter $(\texttt{agent\_id} + \texttt{step}/50) \bmod n$ rotates which 2 of the 10 agents clean at any given time, and a \emph{separate} 200-step zone counter rotates which 5-row band each collector sweeps. The two rotation periods (50 and 200 steps) ensure every agent visits every cleaning slot and every apple zone within one episode, resulting in a structural fairness invariant.

\paragraph{An alternative Sonnet maximin run uses a different mechanism.}
The second Sonnet $\Phi_{\min}$ run took a structurally distinct route. There, $\cR$ never wrote a rotation template into $p$; instead it authored a ``collective threshold'' worked example in which \emph{all} agents synchronously enter a cleaning mode whenever \texttt{waste\_fraction(env)}${\,>\,}0.22$ and a collecting mode when it drops below $0.08$, with no \texttt{agent\_id} phase shift. The resulting synthesizer policy follows this pattern and reaches $\min_i R_i {=} 200$, slightly above the rotation policy of Listing~\ref{lst:min-sonnet}, confirming that the convergence in Table~\ref{tab:strategies} is on the broader \emph{class} (explicit fairness mechanism, 4/4 maximin) rather than on rotation specifically.

\begin{lstlisting}[style=pythonpolicy, caption={Cleanup, $\Phi_{\min}$ -- Sonnet 4.6 (rotation variant; see Appendix~\ref{app:policy-min-sonnet} for the alternative synchronized clean/collect mechanism in the second Sonnet maximin run). Phase-rotated cleaning + zone-rotated collection (verbatim, abridged).}, label={lst:min-sonnet}]
def policy(env, agent_id) -> int:
    if int(env.agent_timeout[agent_id]) > 0:
        return 7  # STAND while removed

    n    = env.n_agents
    step = env._step_count
    wf   = waste_fraction(env)

    # --- ROLE ASSIGNMENT: fair 2-cleaner rotation ---
    # [50-step phases x 20 phases / episode -> each agent cleans 4 phases (~20%)]
    phase        = step // 50
    cleaner_rank = (agent_id + phase) % n   # [cycles 0..9 fairly]
    is_cleaner   = cleaner_rank < 2         # [exactly 2 cleaners per phase]

    # [Emergency override: all agents clean when waste crosses the spawn cliff]
    in_emergency = wf >= 0.40
    if in_emergency: is_cleaner = True

    # [River split: cleaner_slot 0 -> top half, 1 -> bottom half]
    cleaner_slot = (agent_id % 2) if in_emergency else cleaner_rank

    if is_cleaner and wf > 0.04:
        # [Scan 4 orientations from current position; if best beam shot
        #  covers >=3 waste cells, fire; else step to a strictly better neighbor]
        # ... (best_o, best_cnt) = max over 4 directions
        # ... if best_cnt >= 3: rotate to best_o, then CLEAN
        # ... else: scan 4 adjacent cells x 4 orientations for higher yield
        # ... if no waste in beam range: bfs to waste_set in OWN river half,
        #     fall back to whole river

    # --- COLLECTOR LOGIC: 5-zone rotating apple collection ---
    # [5 zones x 200 steps = 1000 steps -> each agent visits every zone once;
    #  2 agents share each zone at any time, reducing competition 10 -> 2]
    bounds     = zone_boundaries(env, 5)
    zone_phase = step // 200
    zone       = (agent_id + zone_phase) % 5
    z_start, z_end = bounds[zone], bounds[zone + 1]

    zone_apples = {(int(env._apple_pos[i][0]), int(env._apple_pos[i][1]))
                   for i in range(env.n_apples) if env.apple_alive[i]
                   and z_start <= int(env._apple_pos[i][0]) < z_end}
    # ... bfs_to_target_set(zone_apples), else bfs_nearest_apple as fallback
    return 7
\end{lstlisting}

\subsection{Gathering (Gemini): wall-aware Voronoi + spatiotemporal targeting}
\label{app:policy-gather}

Best run: $U{=}2.47$, $E{=}0.98$, $\min_i R_i{=}580$. In Gathering, where costs are symmetric, no role differentiation is needed: the entire optimization is spatial and temporal. The synthesizer (i)~partitions cells into BFS-Voronoi territories owned by individual agents, (ii)~runs a single centralized BFS from its own position to score every reachable cell with exact wall-aware distances, and (iii)~for both alive and respawning apples in its territory, computes the \emph{earliest collection time} $\max(\texttt{walk\_dist},\, \texttt{respawn\_timer})$ and targets the minimum. When the agent must wait on a dead apple, it steps onto a \emph{non-spawn} adjacent cell rather than blocking a respawn point. This single sort key  ($\max(\textrm{distance}, \textrm{timer})$)  subsumes both ``go to nearest apple'' and ``camp respawn'' as special cases.

\begin{lstlisting}[style=pythonpolicy, caption={Gathering -- Gemini 3.1 Pro. Voronoi territories + spatiotemporal priority targeting (verbatim, abridged).}, label={lst:gather-gemini}]
def policy(env, agent_id) -> int:
    if int(env.agent_timeout[agent_id]) > 0:
        return 7

    r, c   = int(env.agent_pos[agent_id][0]), int(env.agent_pos[agent_id][1])
    orient = int(env.agent_orient[agent_id])

    # [Wall-aware Voronoi over the gridworld; spawn_points = all apple cells]
    zones        = voronoi_zones(env)
    spawn_points = {(int(p[0]), int(p[1])) for p in env._apple_pos}

    # [Single centralized BFS: exact distance + first move to every reachable cell]
    distances, first_moves = {}, {}
    q = deque([(r, c, 0, None)])
    visited = {(r, c)}
    while q:
        cr, cc, d, fm = q.popleft()
        distances[(cr, cc)]   = d
        first_moves[(cr, cc)] = fm
        for dr, dc in [(-1,0),(1,0),(0,-1),(0,1)]:
            nr, nc = cr + dr, cc + dc
            if 0 <= nr < env.height and 0 <= nc < env.width and not env.walls[nr, nc]:
                if (nr, nc) not in visited:
                    visited.add((nr, nc))
                    q.append((nr, nc, d + 1, fm if fm is not None else (dr, dc)))

    # [Both alive AND respawning apples in MY zone, with their respawn timer]
    my_zone_spawns = []
    for i in range(env.n_apples):
        pr, pc = int(env._apple_pos[i][0]), int(env._apple_pos[i][1])
        if zones.get((pr, pc)) == agent_id and (pr, pc) in distances:
            my_zone_spawns.append((pr, pc, int(env.apple_timer[i]), distances[(pr, pc)]))

    # [SPATIOTEMPORAL PRIORITY: earliest collection time first.
    #  score = max(walk_distance, respawn_timer); ties broken by sooner timer]
    my_zone_spawns.sort(key=lambda x: (max(x[3], x[2]), x[2], x[3]))

    for pr, pc, timer, dist in my_zone_spawns:
        if timer == 0:                  # [Alive apple: walk straight to it]
            if dist == 0: return 7
            fm = first_moves[(pr, pc)]
            return direction_to_action(fm[0], fm[1], orient)
        else:                           # [Dead apple: camp on safe adjacent cell]
            # ... pick nearest neighbor (nr,nc) that is NOT in spawn_points,
            #     navigate there and STAND while waiting for respawn
            ...
    # [Global poaching fallback if zone is empty: nearest alive apple anywhere]
    ...
\end{lstlisting}

\subsection{Cross-condition observations}

Several patterns are visible across these four listings (and confirmed by inspecting the remaining 8 runs not shown):

\begin{itemize}
  \item \emph{Role assignment encodes the welfare objective.} Static \texttt{agent\_id < $\tau$} (Listing~\ref{lst:eff-gemini}) maximizes $U$ but harms $\min_i R_i$. Time-rotated \texttt{(agent\_id + step//$T$) \% n} (Listings~\ref{lst:min-gemini},~\ref{lst:min-sonnet}) maximizes $\min_i R_i$ at $\le 1\%$ efficiency cost (Gemini).
  \item \emph{Convergent rediscovery across LLMs.} Gemini and Sonnet, on independent runs with separate dedicated git branches, both arrive at the \texttt{(id + phase) \% n} duty-rotation idiom with similar phase lengths ($T \approx 50$ steps) in 3/4 maximin runs, and both omit any explicit fairness mechanism under efficiency. The remaining 1/4 maximin run (Sonnet) converges on a structurally distinct synchronized clean/collect mechanism with comparable maximin (Appendix~\ref{app:policy-min-sonnet}). The point of convergence is the \emph{class} of mechanism ($\cR$ writes an explicit fairness mechanism into $p$ under $\Phi_{\min}$, never under $\Phi_U$), not necessarily the rotation idiom itself.
  \item \emph{Game structure dictates strategy class.} Cleanup policies are dominated by role-assignment logic (cleaner vs. gatherer); Gathering policies are dominated by territory partitioning and respawn-timer reasoning, with no role differentiation at all. The researcher does not need to be told which class of strategy to write.
  \item \emph{Earliest-collection-time as a unifying score.} The Gathering policy's $\max(\text{walk}, \text{timer})$ key (Listing~\ref{lst:gather-gemini}) is structurally identical across the 4 Gathering runs (both Gemini and Sonnet), differing only in how the Voronoi diagram is computed, e.g. by Manhattan approximation, fully BFS-based, or cached helper.
\end{itemize}

\end{document}